\documentclass[twocolumn, usenatbib]{aastex62}
\usepackage{graphicx}	
\usepackage{amsmath}	
\usepackage{amssymb}
\usepackage{comment}
\usepackage{color}
\usepackage[title]{appendix}

\newcommand{\zcol}{z_{_{\rm coll}}}
\newcommand{\Rcol}{R_{\rm coll}}
\newcommand{\Rnoz}{R_{\rm noz}}
\newcommand{\Rdiss}{R_{\rm diss}}
\newcommand{\thdiss}{\theta_{\rm diss}}
\newcommand{\IPa}[0]{IP$_{\rm a}$}
\newcommand{\IPb}[0]{IP$_{\rm b}$} 
\newcommand{\DPa}[0]{DP$_{\rm a}$}
\newcommand{\DPb}[0]{DP$_{\rm b}$}
\newcommand{\COa}[0]{CO$_{\rm a}$}
\newcommand{\COb}[0]{CO$_{\rm b}$}

\shorttitle{Non-linear Kink Instability}
\shortauthors{Bromberg, Singh, Davelaar, Philippov}

\begin{document}

\title{Kink instability: evolution and energy dissipation in Relativistic Force-Free Non-Rotating Jets}

\correspondingauthor{Omer Bromberg, Alexander Philippov}
\email{omerbr@tauex.tau.ac.il, sphilippov@flatironinstitute.org}

\author{Omer Bromberg}
\affiliation{The Raymond and Beverly Sackler School of Physics and Astronomy, Tel Aviv University, Tel Aviv 69978, Israel}

\author{Chandra B. Singh}
\affiliation{The Raymond and Beverly Sackler School of Physics and Astronomy, Tel Aviv University, Tel Aviv 69978, Israel}

\author{Jordy Davelaar}
\affiliation{Department of Astrophysics/IMAPP, Radboud University Nijmegen, P.O. Box 9010, 6500 GL Nijmegen, The Netherlands}
\affiliation{Center for Computational Astrophysics, Flatiron Institute, 162 Fifth Avenue, New York, NY 10010, USA}

\author{Alexander A. Philippov}
\affiliation{Center for Computational Astrophysics, Flatiron Institute, 162 Fifth Avenue, New York, NY 10010, USA}
\affiliation{Moscow Institute of Physics and Technology, Dolgoprudny, Institutsky per. 9, Moscow region, 141700, Russia}

\begin{abstract}
We study the evolution of kink instability in a force-free, non-rotating plasma column of high magnetization.
The main dissipation mechanism is identified as reconnection of magnetic field-lines with 
various intersection angles, driven by the compression of the growing kink lobes.
We measure dissipation rates ${\rm d} U_{B\phi}/{{\rm d}t} \approx -0.1 U_{B\phi}/\tau$, where $\tau$
is the linear growth time of the kink instability. This value is consistent with the expansion velocity of the kink mode, which drives the reconnection. The relaxed state is close to a force-free Taylor state. We constraint the energy of that state using considerations from linear stability analysis.  
Our results are important for understanding magnetic field dissipation in various extreme
astrophysical objects, most notably in relativistic jets. We outline the evolution of the kink instability in such jets and derive constrains on the conditions that allow for the kink instability to grow in these systems. 
\end{abstract}


\section{Introduction}

Relativistic jets power some of the most luminous astrophysical objects we know, like gamma-ray bursts (GRBs), microquasars and radio loud galaxies (RLG). It is generally accepted that the jets are launched electromagnetically, most likely by the winding of magnetic field lines that thread a rotating compact object \citep{bz77,kom01}. The winding generates Poynting-flux at the expense of rotational energy, which is later collimated to form a jet. Though the process of magnetic jet launching seems to be well understood, the jet physics at large distances is still a matter of active debate \citep[e.g. see a review by][]{2015SSRv..191..441H}. One of the most fundamental questions is where and how jets dissipate their magnetic energy. This has important implications on particle acceleration and emission mechanisms in the jets, the fraction of magnetic energy carried by the jets at large distances, and on the jets stability properties. 

The theory of magnetic jets stability was originally developed for magnetic confinement of plasma in Tokamak facilities \citep[e.g.][]{1973PhFl...16.1909F,1973PhFl...16.1894R,1975SvJPP...1..389K}. This theory was later applied to astrophysical jets where analytic and numerical studies were conducted in non-relativistic \citep[e.g.][]{1979SoPh...64..303H,1996A&A...314..995A} as well as highly relativistic regimes \citep[e.g.][]{1998ApJ...493..291B,1999MNRAS.308.1006L}. In toroidal-field dominated jets, the fastest growing instability is known as kink instability. This current-driven instability (CDI) generates helical deformations in the jet, which can lead to an efficient dissipation of the jet's magnetic energy and may even disrupt the jet altogether. Linear stability analysis by \citet{1999MNRAS.308.1006L} and by \citet{2000A&A...355..818A} found the growth rates and typical wavelengths of the instability. Later \citet{2000A&A...355.1201L} showed that the non-linear state is well characterized by a fastest growing mode as predicted by the linear stability analysis. The basic results of these studies, mainly the growth rates, were confirmed with numerical MHD simulations \citep[e.g.][]{2009ApJ...700..684M,2012ApJ...757...16M}). However, a detailed numerical study of the non-linear evolution of the instability in the relativistic regime, the relaxation condition and, most importantly, the amount and rate of the magnetic energy dissipation was not performed.   

In this work we conduct a systematic study of the evolution of kink instability in highly magnetized, initially force-free columns, using relativistic magneto-hydrodynamic (MHD) simulations. 
We start by summarizing the linear theory of kink instability in various magnetic field configurations in Section 2. We then describe the non-linear evolution of the kink mode and outline the predictions for the magnetic relaxation, which were established in the low-magnetization regime 
(Section 3). In Section 4 we discuss the minimal energy state and how it can be used to predict the amount of dissipated energy. In Section 5 we outline the numerical setup, and in Section 6  we report our findings. We identify the dissipation mechanism, verify the relaxation criterion and quantify the amount of energy dissipation that takes place in the process. We discuss the implications for astrophysical jets and twisted magnetic loops (Section 7) and conclude in Section 8.

\section{Kink Instability linear evolution}
CDI modes tend to grow on resonant surfaces which satisfy the condition $\bf{k}\cdot\bf{B}=0$, where $\bf{k}$ is the wave vector of the growing mode \citep{1973PhFl...16.1894R,1975SvJPP...1..389K}, and {\bf B} is the vector of the local magnetic field. In cylindrical geometry this translates to the condition
$kB_z+(m/r)B_\phi=0$ with $k,m$ being the wave numbers in the longitudinal and azimuthal  
directions respectively, and we use standard cylindrical coordinates $(r,\phi,z)$. In a periodic box, the vertical wavenumber can be expressed as $k=2\pi n/ L $, where $n$ is an integer number and $L$ is the longitudinal box size. The resonant condition can also be written as 
\begin{equation}
    kP+m=0,
\end{equation}
where $P\equiv rB_z/B_\phi$ is the pitch of the magnetic field.
Linear stability analysis for jets of finite length show that resonant modes grow on discrete surfaces which fulfill the condition $k_{\rm res}\simeq -m/P_0$, where $P_0$ is the pitch at the axis\footnote{In a magnetic configuration of a uniform pitch, $P_0$ is equal to the radius of jet core, which carries most of the current that supports the toroidal-magnetic-field component.}. The fastest growing mode is the $m=-1$ mode, known as the kink mode. In practice, it grows over a range of wavelengths, where the maximum growth rate occurs at a wave number
\begin{equation}\label{eq:kmax}
    k_{\rm max}\simeq 0.745\times 1/P_0,
\end{equation} 
having a growth rate  
\begin{equation}\label{eq:Lambda_max}
 \Lambda_{\rm max}=0.133v_A/P_0, 
\end{equation}
where $v_A$ is the Alfv\'en velocity. These scalings are  almost independent of the pitch profile \citep{2000A&A...355..818A}.

\section{Non-linear evolution and relaxation}\label{sec:theory_relax}

Though the linear growth of the kink mode has a rather weak dependence on the pitch profile, its evolution in the non-linear regime changes with the pitch profile. A theoretical understanding of the non-linear regime and of the relaxation process was obtained for non-relativistic configurations \citep{1975SvJPP...1..389K}. The plasma in this case has low magnetization, $\sigma=B^2/4\pi\rho c^2\ll 1$, and it resides in a periodic box with length $L=2\pi P_0$ along the jet axis, which only allows for $n=1$ mode to grow. 

\begin{enumerate}
    \item In configurations where $P(r)$ increases with $r$ (increasing pitch, IP) there is a resonant surface which corresponds to a fastest growing mode,  $P(r_{\rm res})=1/k_{\rm max}\approx 4P_0/3$. The mode's wavelength can be expressed as 
    \begin{equation}
        \lambda_{\rm res}=\frac{2\pi}{k_{\rm res}}\simeq\frac{8\pi}{3}P_0.
    \end{equation}
    It generates a helical twist in the jet, which grows inwards to the resonant surface, and leads to the formation of a large-scale current sheet at a radius $\sim r_{\rm res}$. As the mode continues to grow, the current sheet extends to the regions between the kink lobes, gets compressed and eventually breaks due to resistive instabilities \citep{1975SvJPP...1..389K}. The dissipation may proceed in a more  stochastic fashion, through small scale current sheets or turbulence, while maintaining the global helical shape of the kink mode.  
    
    Since in this case the kink mode perturbs only the jet inside of $\sim r_{\rm res}$ it is termed an \emph{internal kink} mode. If the resonant surface is located outside the boundary of the current carrying core (e.g. the jet boundary), the mode will spread out until it will engulf the entire core, generating a global helical structure. We term this mode an \emph{external kink} mode\footnote{
    In plasma physics literature a kink mode is called external if it grows on the plasma-vacuum boundary, which is absent in astrophysical systems. Instead, we term a kink mode that grows on, or outside the boundary of the current-carrying core and deforms it's shape as an external mode}.
    
    \item In configurations where the pitch profile decreases with radius (decreasing pitch, DP) there is no surface fulfilling the resonance condition in eq. \ref{eq:kmax}. Any large scale kink mode that grows is expected to break apart, avoiding the formation of a prominent global current  sheet. The result is a more stochastic evolution, likely without a large scale helical pattern.
    As the global mode continues to grow, it may eventually disrupt the entire jet. Magnetic configurations with the DP profile are naturally more unstable, so it remains questionable if and how these configurations can be realized in the first place.
    \item The case of a constant pitch can be considered as a special case. Since the fastest growing mode corresponds to a pitch value $P(r_{\rm {res}})=1/k_{\rm max}>1/P_0$, there is no resonant surface. In the limit of a small pitch ($P_0\ll L, R_j$) the evolution will be similar to the DP case. A non-resonant mode will grow at $r\simeq P_0$, and will spread outward leading to a global/stochastic dissipation. In the limit of a large pitch, the large value of $P_0$ stabilizes it against the growth of internal modes. Moreover, the kink growth rate quickly decreases with growing $P_0$, thus, the jet becomes stable for further dissipation by the kink instability \citep{2000A&A...355..818A}. 
 \end{enumerate}

The dissipation of electromagnetic (EM) energy takes place mostly during the non-linear stage through reconnection and stochastic/turbulent dissipation. Although the magnetic configuration changes during the dissipation process, 
the total helicity is roughly conserved 
\citep[e.g.][]{1974PhRvL..33.1139T,1986RvMP...58..741T,2000PhPl....7.1623T}.
The magnetic field configuration gradually relaxes into a minimal energy state which is known as a Taylor state \citep{1974PhRvL..33.1139T}, which maintains 
\begin{equation}
    {\bf j}(r) = \alpha {\bf B}(r), 
\label{eq.taylor}
\end{equation}
where $\alpha$ is constant.
It can be expressed as
\begin{equation}\label{eq:alpha}
    \alpha=\frac{{\bf B}\cdot\left({\bf\nabla}\times{\bf B}\right)}{B^2}.
\end{equation}
Note that eq. \ref{eq.taylor} corresponds to a force-free state, since ${\rm {\bf j}}\times{\rm{\bf B}}=0$. A magnetic field configuration which is both cylindrically symmetric and obeys condition (\ref{eq.taylor}) can be expressed as:
\begin{eqnarray}\label{eq:B_bessel}
B_z &=& B_0 J_0(r\alpha)\nonumber \\
B_\phi &=& B_0 J_1(r\alpha)
\end{eqnarray}
where $J_0$ and $J_1$ are the zeroth and first Bessel functions of the first kind. 
This configuration is unstable for $m=-1$ kink modes that  satisfy \citep{1962PhRv..128.2016V}
\begin{equation}\label{eq:alpha_stable}
    k<0.272\alpha.
\end{equation}
In a periodic box of length $L$ the minimal $k$ that can be excited is $k=\pi/L$, which corresponds to a wavelength $\lambda=2L$. 
Thus, if $\pi/L > 0.272\alpha$, the configuration is stable to kink. Namely, for a given box size configurations with $\alpha \lesssim 4\pi/L$ are stable for kinking. Note that in the stable case the value of the pitch at the axis is
\begin{equation}\label{eq.KS}
    P_0 = \frac{2}{\alpha}\gtrsim\frac{L}{2\pi}, 
\end{equation}
which is just the Kruskal-Shafranov (KS) criterion \citep{1956AtEnerg.5...38,1958RSPSA.245..222K} for the stabilization of kink instability. Jets with a Bessel profile and a small aspect ratio ("infinitely long jets") are stable for all kink modes if they satisfy
\begin{equation}\label{eq:alphaRj}
\alpha R_j\leq3.176,    
\end{equation}
where $R_j$ is the cylindrical jet radius \citep{1962PhRv..128.2016V}. For values of $\alpha R_j$ in the range $3.176\leq\alpha R_j\leq3.832$ the jets become increasingly unstable until for $\alpha R_j > 3.832$ they are unstable for all modes with $k<0.272\alpha$. It is important to notice that for $\alpha R_j\leq3.832$ the first zero of $J_0$ is located inside $R_j$, and the first zero of $J_1$ falls outside $R_j$. 
This implies that $B_z$ flips its sign in the outer part of the jet, while $B_\varphi$  maintains its direction. We find evidence for such a behavior in several configurations that we tested in this work. All of them reached a condition of marginal stability with $\alpha R_j$ being close to the value given by eq. \ref{eq:alphaRj}.

\section{Minimal energy state}\label{sec:Emin}
If the final configuration is fully relaxed, the magnetic field profile can be described by the set of Bessel functions given in eq. \ref{eq:B_bessel}.  Three parameters are required to calculate the final EM energy in this case: $B_0, \alpha$ and $R_j$. The dissipation process conserves two quantities to a good accuracy: the total helicity and the total axial magnetic flux. 
A third condition can come comes from constraining the final $\alpha$, (e.g. 
by the stability criterion given in eq \ref{eq:alphaRj}), or the radius of the dissipated region.

The total helicity in a volume is defined by
\begin{equation}
H\equiv\int_V{\rm\bf A}\cdot{\rm\bf B}~dV.    
\end{equation}
As such, it is a gauge-dependent quantity. Gauge invariance is possible in specific magnetic field typologies, for example when ${\rm\bf B}$ is tangent to the boundary of the volume, and its evolution conserves longitudinal magnetic flux \citep{Browning2008}. The situation of an axisymmetric field with vanishing radial component of the magnetic field on the boundary is ideal for the helicity conservation. In this case the helicity can be described as:
\begin{equation}\label{eq:H}
    H=2\pi L\left[2\int_0^{R}\frac{\Psi(r')}{2\pi}\frac{2I(r')}{r'}dr'+\left.\left(A_z(r)\frac{\Psi(r)}{2\pi}\right)\right|_0^R\right],
\end{equation}
where $\Psi(r)$ is the magnetic flux within radius $r$, and $I(r)$ is the current within the same radius. Taking a gauge $A_z(R)=0$, the second term vanishes and we are left with the first one, which we identify as
\begin{equation}\label{eq:KH}
    K(R)\equiv 2\int_0^R\frac{\Psi(r')}{2\pi}\frac{2I(r')}{r'}dr'.
\end{equation} 
$K(R_j)$ is largely conserved throughout the evolution of the system. 
If magnetic configuration in the final state can be approximated as a cylindrically symmetric Taylor state, $K$ and $\Psi$ can be expressed as (see Appendix A):
\begin{align}
    K&=\frac{B_0^2}{\alpha^2}\Upsilon(R_j),\label{eq:KB}\\
    \Psi&=2\pi B_0\int_0^{R_j}J_0(\alpha r)rdr,\label{eq:PsiB}
\end{align}
with
\begin{equation}
\small{
\Upsilon(R_j)=\int_0^{\alpha R_j}\left[J_0(\xi)^2+\newline
    J_1(\xi)^2\right]d\xi-J_0(\alpha R_j)J_1(\alpha R_j){R_j}}.     
\end{equation}
Substituting the initial $K$ and $\Psi$ values in 
eqs. \ref{eq:KB} and \ref{eq:PsiB}, and adding a constraint on the relaxed configuration, for example eq. \ref{eq:alphaRj}, gives a closed set of equations from which we can estimate the energy in the final state. 

Though the outlined theory of magnetic relaxation has been applied in the non-relativistic regime applicable for solar flare \citep{Browning2008}, it is unclear whether the same theory applies to relativistic plasma in extreme astrophysical environments of relativistic jets or twisted magnetic loops in the accretion disk coronae. First, since $\sigma\gg1$, the dissipation process generates thermal pressure, which can be of the order of the mass energy density of the plasma, and  it is unclear whether a force-free condition can be sustained. Second, it is unclear which process, turbulence or reconnection, dominates the dissipation process.
We employ numerical simulations to test the non-linear evolution of kink instability in relativistic plasma and compare the results to the expectations from the non-relativistic theory.

\section{Numerical setup}
For our studies of the kink instability we use the {\it Software: }{{\tt PLUTO}\citep{2007ApJS..170..228M,2012ApJS..198....7M}}, a three-dimensional relativistic MHD code designed to simulate astrophysical flows with high Mach numbers and moderate to high values of the magnetization parameter. 
\citep[e.g.][]{2010MNRAS.402....7M,2013MNRAS.436.1102M,2013MNRAS.434.3030B}. {\tt PLUTO} has a very flexible numerical scheme, which allows to test how the details of the implementation affect the results. Our chosen scheme consists of a third order Runge-Kutta time stepping, piecewise parabolic reconstruction with harmonic limiter, HLL Riemann solver, and we use a Courant number of 0.3. In the case of high $\sigma$, low plasma $\beta$ regime more accurate solvers like HLLD can lead to numerical problems \citep{2007ApJS..170..228M,2014MNRAS.442.2228A}. In order to avoid unphysical states, slope-limited reconstruction with the MinMod limiter is adopted to handle shocks, and we use constrained transport to enforce ${\rm div}{\bf B} =0$. We use an ideal equation of state with an adiabatic index 4/3. 

To examine the evolution of \emph{internal kink} modes we set the computational box inside the jet, so that the jet boundaries lie outside the box. Our study is focused on relativistic jets, however, the approximations we make are relevant also for other systems, such as twisted magnetic loops in the accretion disk coronae.
We therefore set up a second configuration where a high $\sigma$ core (the "loop") is embedded in a magnetized external medium, which is relevant for such a case  \citep[e.g.][]{2011ApJ...729..101G}. In such configuration we examine the evolution of \emph{external kink}. 
We use a Cartesian grid with periodic boundaries in the direction of the jet axis, $z$, and outflow boundary conditions in the transverse, $x$-$y$, directions.

We perform simulations in a reference frame comoving with the jet. In the coronal configuration, this setup corresponds to the frame of the magnetic loop. For simplicity we neglect gradients in the longitudinal velocity and rotation \citep{2009ApJ...700..684M}. Even though the magnetic field in the jet is predominantly toroidal in the lab frame, the poloidal field cannot in general be neglected because one has to compare the fields in the comoving frame where the toroidal field is much lower. For example, in equilibrium configuration with cylindrical symmetry, the poloidal and toroidal fields in the comoving frame are comparable \citep{lyub09}. Therefore, analysis of kink instability for astrophysical jets has to take into account the non-negligible poloidal field. In addition, since it is very likely that before the flow becomes kink-unstable the plasma is cold, force-free configuration is a good initial condition. In the absence of rotation\footnote{In the presence of rotation, the hoop stress can be entirely compensated by the electric force. In  this case cylindrically symmetric configuration is known to be stable to kink instability
\citep{1996MNRAS.281....1I,1999MNRAS.308.1006L}.
However, if the profile of poloidal field  shows substantial transverse gradient, growth of the instability in rotating and non-rotating equilibria is qualitatively similar \citep{2017MNRAS.468.4635S}. A numerical investigation of this case will be performed in a separate work.} the hoop stress has to be balanced by the gradient in the poloidal magnetic pressure, so an equilibrium configuration generally has a core of poloidal field near the axis.

\begin{figure*}
    \hspace*{-2cm}
    \begin{center}
    \includegraphics[width=0.32\textwidth]{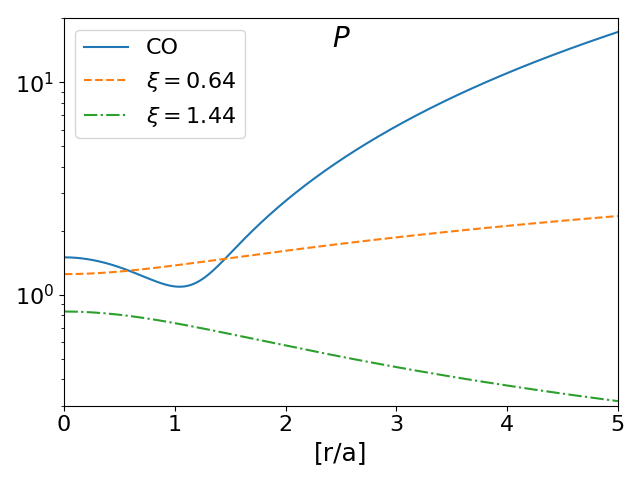}
    \includegraphics[width=0.32\textwidth]{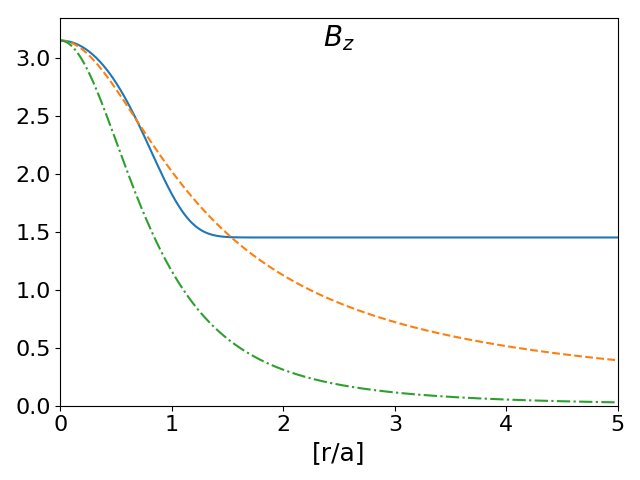}
    \includegraphics[width=0.32\textwidth]{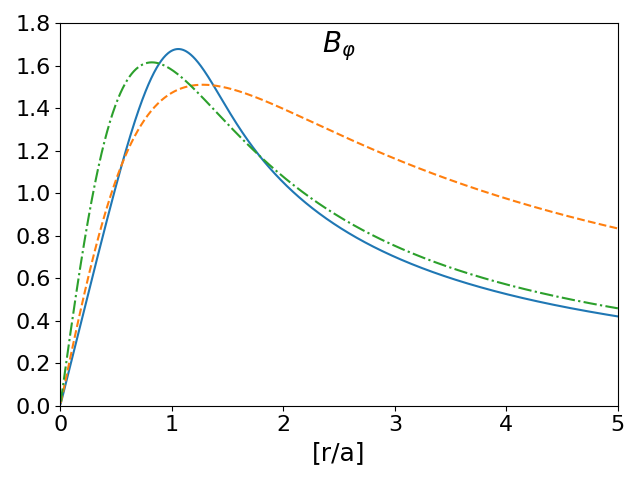}
    \includegraphics[width=0.32\textwidth]{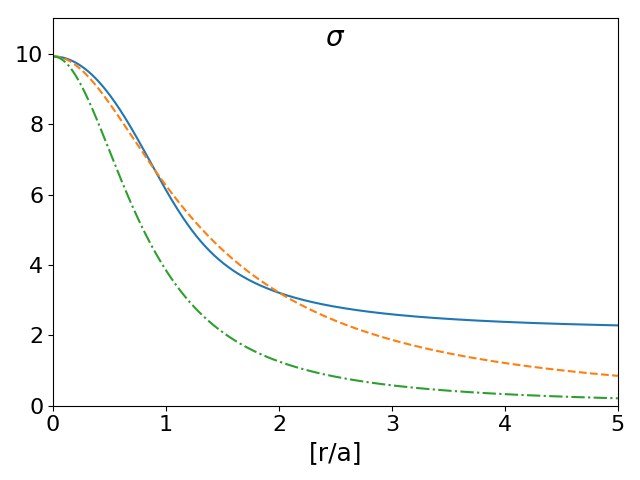}
    \includegraphics[width=0.32\textwidth]{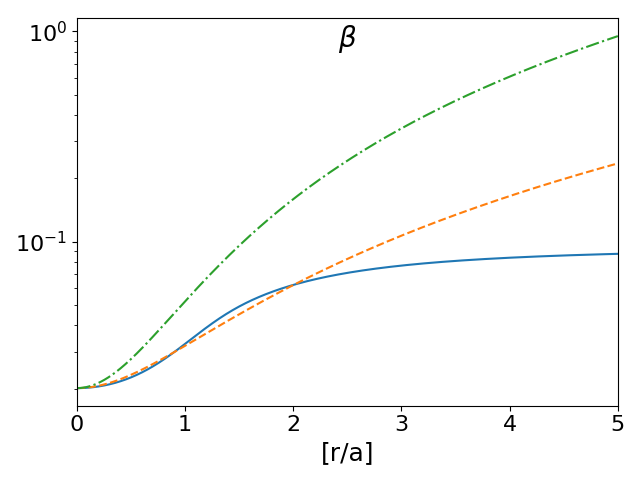}
    \end{center}

    \caption{The initial configuration of the three profiles tested in this work Coronal (CO,blue), increasing pitch (IP, dashed orange) and decreasing pitch (DP, dash dot green). Panels show from left to right, top to bottom: Pitch (in log scale), $B_z$, $B_\varphi$, $\sigma$ and plasma $\beta$ (in log scale). }
    \label{fig:init_profiles}
\end{figure*}

We set up a helical magnetic field with a non-rotating, force-free configuration \citep{2009ApJ...700..684M}:
\begin{eqnarray}
B_{\rm r}(r) &=& 0\nonumber,\\
B_{\rm z}(r) &=& \frac{B_0}{\left[1+(r/a)^2\right]^\zeta},\\
B_{\phi}(r) &=& \frac{aB_z}{r}\sqrt{\frac{\left[1+(r/a)^2\right]^{2\zeta}-1-2\zeta(r/a)^2}{2\zeta-1}}\nonumber,
\end{eqnarray}
This profile has a monotonic pitch determined by the parameter $\zeta$. The pitch is increasing for $\zeta<1$ and decreasing if $\zeta>1$. The radius of the $B_z$ dominated core is of the order of the value of pitch at the axis, $P_0=a\sqrt{1/\zeta}$. 
We consider two values of $\zeta=0.64,1.44$ representing configurations of IP and DP with $P_0=1.25a,5/6a$ respectively. A third configuration we study is based on \citet{2013MNRAS.434.3030B}, which is also a force-free and static configuration. In this case the helical core is embedded in a uniform axial "external" field.
Such configuration can be applicable for twisted magnetic field loops in  accretion disk coronae or in magnetospheres of magnetars \citep{2009ApJ...703.1044B,2013ApJ...774...92P}. The field configuration has the form:
\begin{eqnarray}
B_\phi &=& \frac{B_0 R}{r}\sqrt{1-e^{^{-\frac{r^4}{a^4}}}}\nonumber,\\
B_z &=& \frac{B_0RP_0}{a^2}\sqrt{1-\sqrt{\pi}\left(\frac{a^2}{P_0^2}\right){\rm erf}\left(\frac{r^2}{a^2}\right)},
\end{eqnarray}
where $R$ is the cylindrical radius of the computational box, $P_0$ is the value of the pitch at the axis that sets the relative strength of two field components.
In this work we study the kink evolution in the case of initial high magnetization at the axis, defined as $\sigma\equiv b^2/4\pi\rho c^2$. We perform simulations with a peak magnetization $\sigma=10$, and set a uniform pressure and mass density in the box as an initial condition. We normalize all length scales  by $a$, time units by $a/c$, energy density by $\rho c^2$ and strength of the magnetic field by $\sqrt{4\pi\rho_0c^2}$. In these units the values of the gas density and pressure are  $\rho=1$ and $p=0.01$ respectively.
The magnetic fields and related pitch profiles used in this work are presented in Fig. \ref{fig:init_profiles}.
To initiate the kink instability we introduce random perturbations to the radial velocity $v_r = \eta_{_N}\delta v e^{-r/2a}$, 
where $\delta v = 0.1c$ and $\eta_{_N}$ is a random number drawn from a uniform distribution in the range $\{-1,1\}$. We also performed simulations with $\delta v = 0.01c$ and found no difference in the linear growth rates and the non-linear evolution.

We set the size of the box in the longitudinal direction so that it fits several kink wavelengths ($L\simeq2\pi n/k_{\rm max}$, $n>1$). This allows us to test the effect of interactions of multiple modes on the dissipation process. 
To study the dependence of the dissipation on the number of excited modes, we vary the size of the computational box, thus allowing for different number of kink wavelengths to grow. Table \ref{tab:res} summarizes the magnetic profiles studied in this work and the box sizes we used.
To sample the dissipation scales properly in the MHD simulations we need to resolve the core with at least 15 cells per unit radius $a$. A convergence tests with 30 and 45 cells per unit radius showed no significant difference in the evolution of the kink instability. The convergence tests are presented in Appendix B.

\begin{table}
\caption{Simulations parameters}
\begin{center}
\begin{tabular}{|c|c|c|c|c|} 
 \hline
 Name & $\sigma_0$ & Box & Resolution & $t_f$  \\ 
 &&Dimension&&$[a/c]$\\ 
 \hline
 \IPa& 10 & $80\times80\times20$ & $1200\times1200\times300$&2000\\ 
 \IPb& 10 & $120\times120\times40$ & $1800\times1800\times600$&3000\\ 
 \hline
 \DPa & 10 & $80\times80\times14$ & $1200\times1200\times210$&1760\\ 
 \DPb & 10 & $100\times100\times28$ & $1500\times1500\times420$&1760\\ 
 \hline
 \COa & 10 & $30\times30\times20$  & $450\times450\times300$&2000\\ 
 \COb & 10 & $30\times30\times80$ & $450\times450\times300$&2000\\ 
 \hline
\end{tabular}\label{tab:res}
\end{center}
\end{table}

\section{Results}
\subsection{Overall structure and growth rates} 
\begin{figure*}
    \centering
    \includegraphics[width=0.75\linewidth]{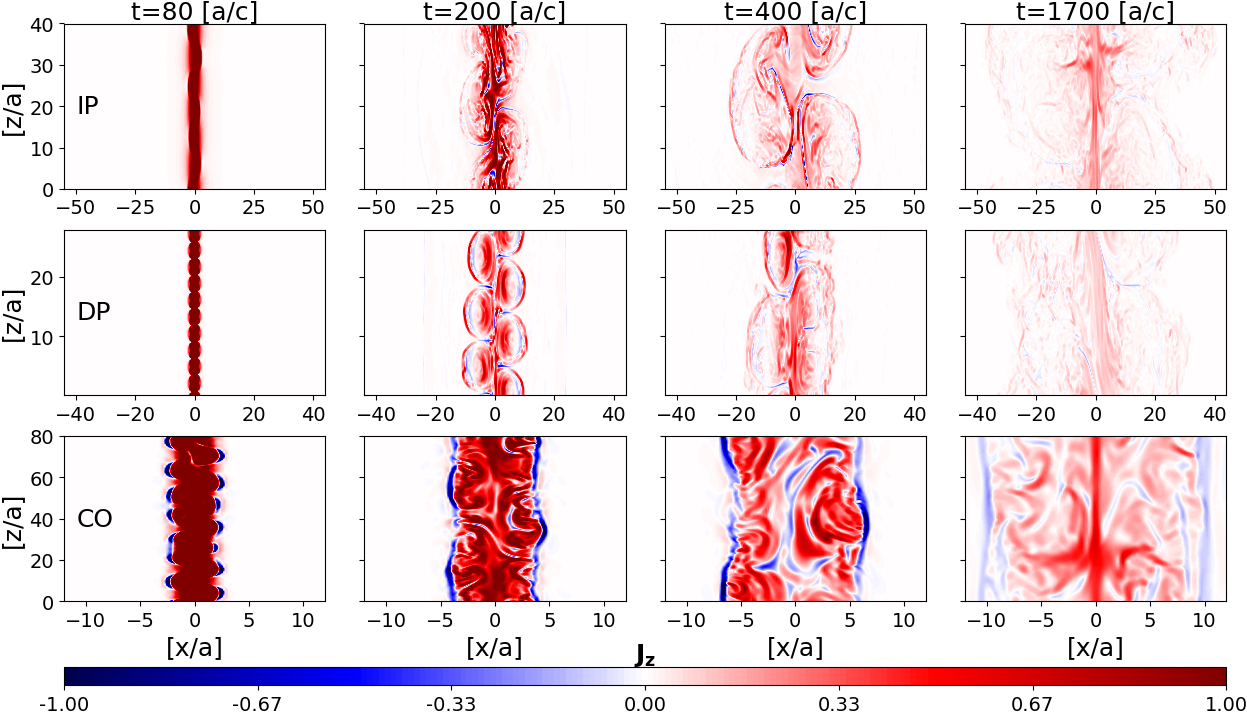}
    \caption{The evolution of the kink instability in cases \IPb{}, \DPb{} and \COb{}. Shown are  values of $J_z$ on the $x$-$z$ plane. Current sheets are seen as peaked color filaments.}
    \label{fig:Jz_Mosice}
\end{figure*}

\begin{figure*}
    \centering
    \includegraphics[width=0.75\linewidth]{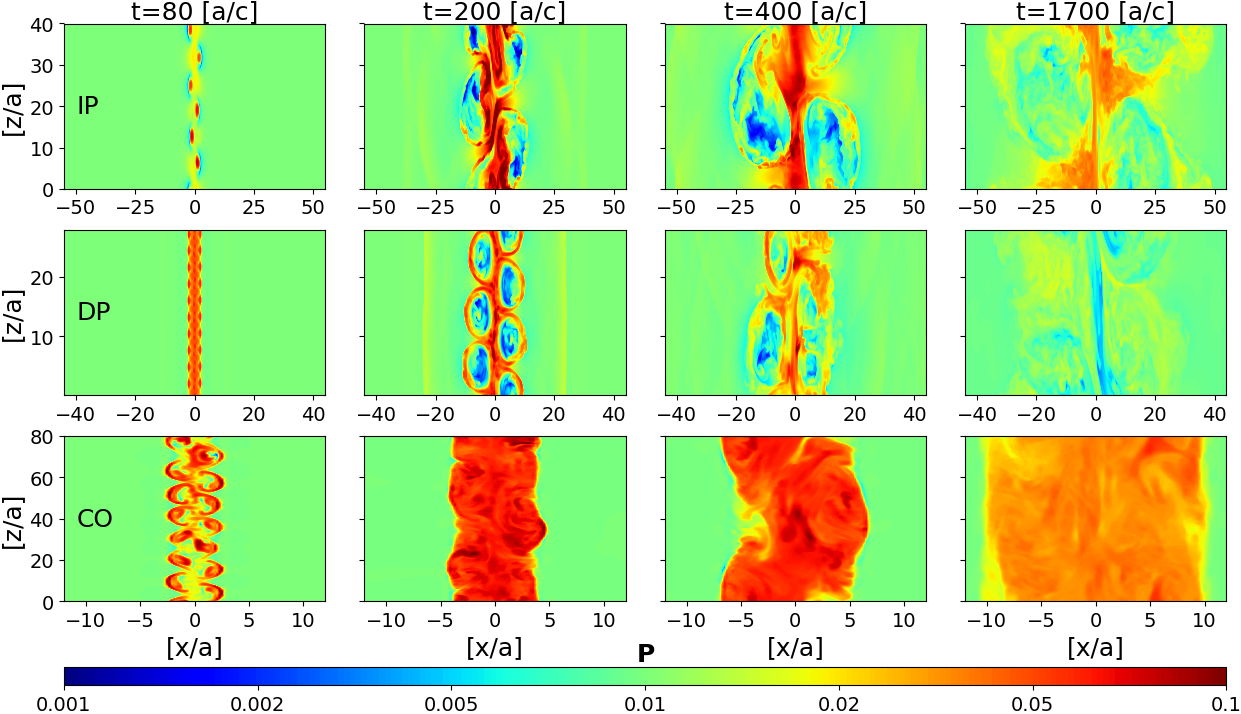}
    \caption{Same as fig. \ref{fig:Jz_Mosice} for the thermal pressure, shown in logarithmic scale. Regions of high pressure match the peak filaments in $J_z$, implying that most of the dissipation is occurring in current sheets. The pressure in the right most column is in the course of becoming evenly distributed across the dissipated region.}
    \label{fig:P_Mosice}
\end{figure*}

\begin{figure*}
    \centering
    \includegraphics[width=0.75\linewidth]{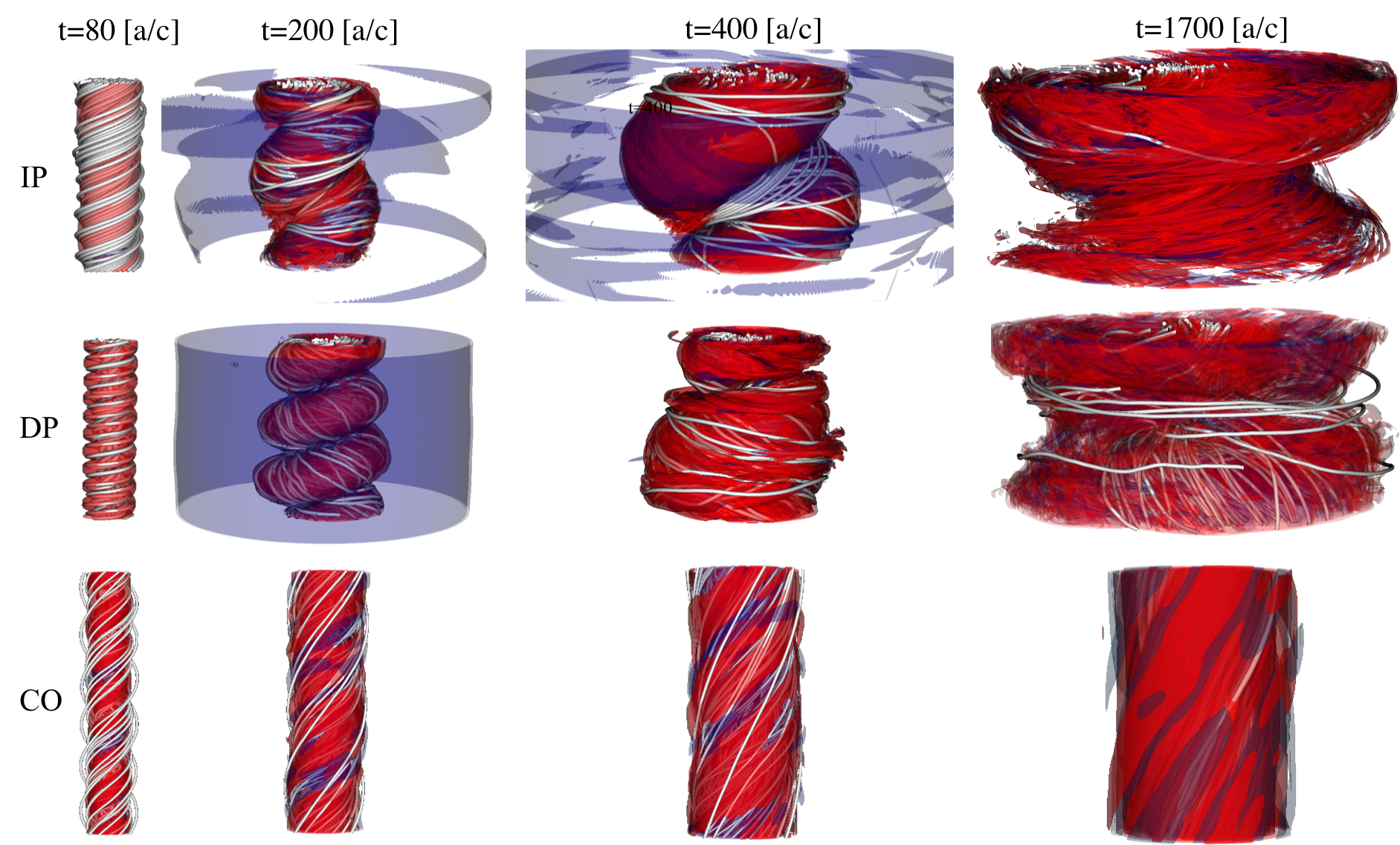}
    \caption{3D color rendering of $J_z$ at the same times and color range as in fig. \ref{fig:Jz_Mosice}. Magnetic field lines are shown as white tubes}
    \label{fig:Jz_Mosice_3D}
\end{figure*}

\begin{figure}
    \centering
    \includegraphics[width=1.05\linewidth]{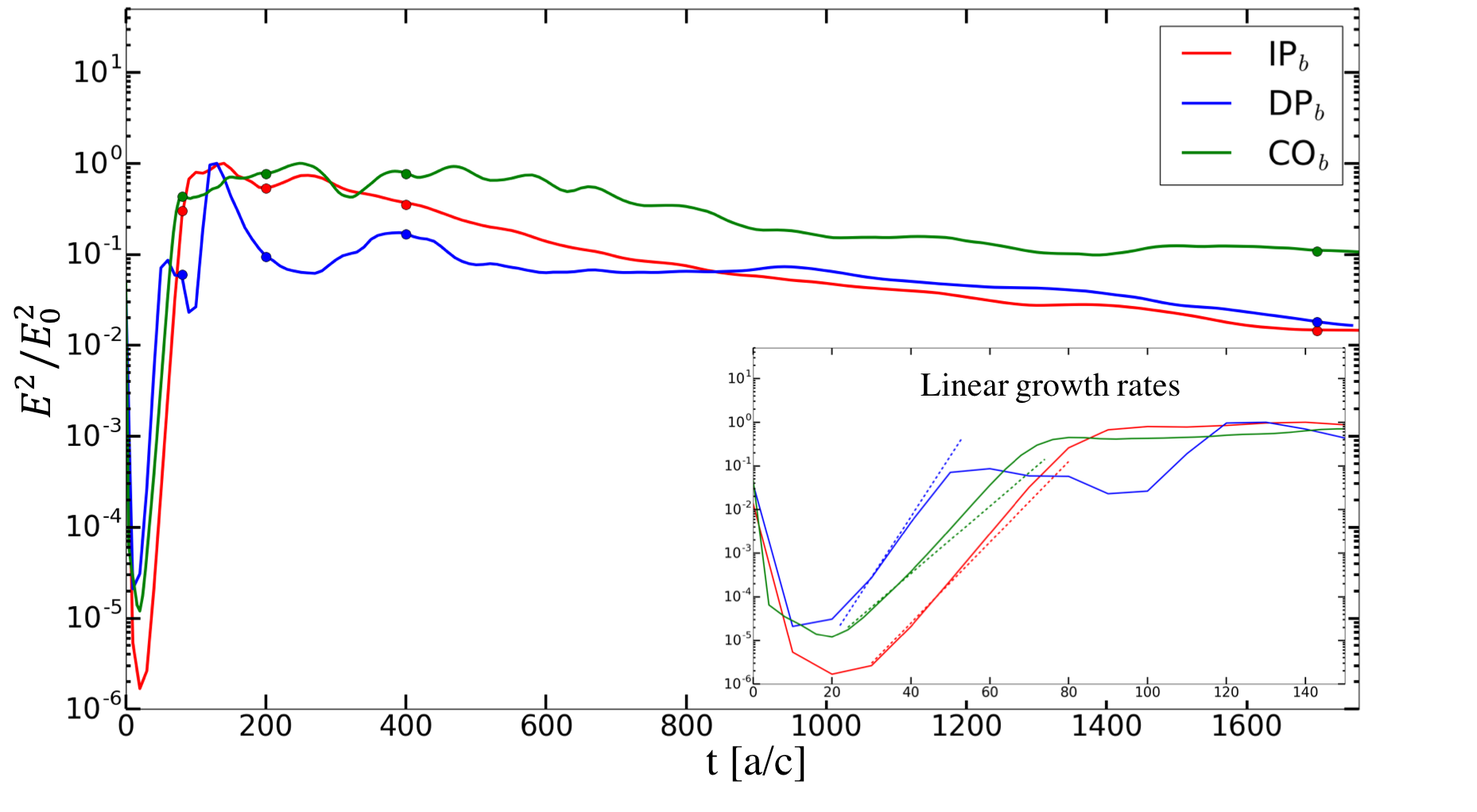}
    \caption{Evolution of the kink mode shown as the electric energy in the three simulated profiles: \IPb, \DPb, and \COb. Three phases are evident: i) linear growth; ii) mode inverse cascade; iii) turbulence phase. The filled circles on the three curves, mark the times at which the snapshots in figs. \ref{fig:Jz_Mosice}- \ref{fig:Jz_Mosice_3D} are taken.
    Comparisons to the theoretical linear growth rates of $k_{\rm max}$ (eq. \ref{eq:Lambda_max}) are shown as dotted lines plotted over the growth curves of $E^2$ in the subplot at the bottom right corner. In all three models the growth rates are in good agreement with the theoretical predictions.}
    \label{fig:Growth_rate}
\end{figure}

The evolution of the kink instability can be characterized by several stages, depicted in figs. \ref{fig:Jz_Mosice} - \ref{fig:Jz_Mosice_3D}. The figures show a series of snapshots from various evolutionary stages of the studied systems. We show results for the large box runs, cases \IPb{} (top), \DPb{} (middle) and \COb{} (bottom). Figs \ref{fig:Jz_Mosice} and \ref{fig:Jz_Mosice_3D} show the current density in the $z$ direction and fig. \ref{fig:P_Mosice} shows the thermal pressure at the same times. The growth rates of the kink modes in the three cases are seen in fig. \ref{fig:Growth_rate}, which shows the average value of $E^2$ in the box normalized by the initial value. 
The evolution in all cases is 
characterized by an initial fast exponential rise, evident as a linear growth in fig. \ref{fig:Growth_rate} and demonstrated in the left most panels of \ref{fig:Jz_Mosice} - \ref{fig:Jz_Mosice_3D}. The growth rates in this stage match the analytic predictions of the linear theory quite well, as can be seen in the zoomed-in box of fig. \ref{fig:Growth_rate}. 

Beyond the linear stage, the evolution depends on the magnetic field configuration, in particular on the pitch profile. 
In the IP case the kink mode grows on a resonant surface. Since the mode is mostly confined to that surface it grows faster in the longitudinal direction, increasing the width of the individual kink lobes until they touch each other (see fig. \ref{fig:Jz_Mosice} at $t=200$ [a/c]).
At this point the exponential growth saturates and the kink mode starts to "inverse cascade" through a series of coalescence events (mergers), where in each merger the longitudinal wave number, $n$, is reduced by unity. This phase is seen in fig. \ref{fig:Growth_rate} as a series of bumps in the value of $E^2$. 
As the inverse cascade continues the longitudinal wave vector, $k$ decreases until it  reaches  $k_{min}=\frac{\pi}{L}$, corresponding to an $n=1/2$ wave number, and the merger stops (the right most image in figs. \ref{fig:Jz_Mosice}-\ref{fig:Jz_Mosice_3D}).
The merger
process breaks the ordered structure of the magnetic field, forming a stochastic turbulent configuration, 
which slowly relax to a minimal energy state once the mergers ends (after $\sim t=400$ [a/c]). 

In the DP case, there is no resonant surface. The kink mode grows in amplitude as well as in width, until the kink lobes touch each other and begin to merge.
Here we identify a major merging episode, which brings the wave number down to a low $n$ value in a single event (as oppose to the gradual merging process in the IP case). 
It is followed by secondary, weaker events, which destroy the structure of kink mode completely. 
As in the IP case, the merger process breaks the global structure of magnetic field forming a stochastic turbulent structure which eventually relaxes to a stable configuration. 

In the CO case the resonant surface is located very close to the edge of the helical core. As a result, the kink mode grows close to the core edge and quickly becomes an \emph{external mode} to the core. The high magnetic tension of the external longitudinal field prevents the kink mode on the core boundary from growing to a large amplitude with respect to the core's cross sectional radius. Instead, the growth takes place mostly along the boundary, more extremely than in the IP case, creating a current sheet at edge of the core which quickly breaks down. As the instability continues to grow, the kink mode inverse cascades to longer wavelength via a series of mergers that ends at $600\lesssim t\lesssim800$ [a/c] when it reaches the smallest $k$ allowed in the box. It results in a 
mildly perturbed core with a stochastic structure of magnetic fields, which slowly relax to the minimal energy state. During the evolution of the kink instability, the radius of the dissipated core grows. As the core pushes against the magnetic field in the medium, external matter mixes into the core through instabilities at the boundary.


\subsection{Energy dissipation}
The dissipation of the EM energy occurs mostly in current sheets and is tightly related to the evolution of the kink instability. The current sheets are evident in fig. \ref{fig:Jz_Mosice} as local extrema in the current density with filamentary shape. Fig. \ref{fig:P_Mosice} shows the corresponding thermal pressure measured at the same time. The pressure peaks match the location of the filaments of $J_z$, indicating that most of the dissipation occurs in the current sheets.

\begin{figure}
    \centering
    \hspace*{-15pt}
    \includegraphics[width=1.05\linewidth]{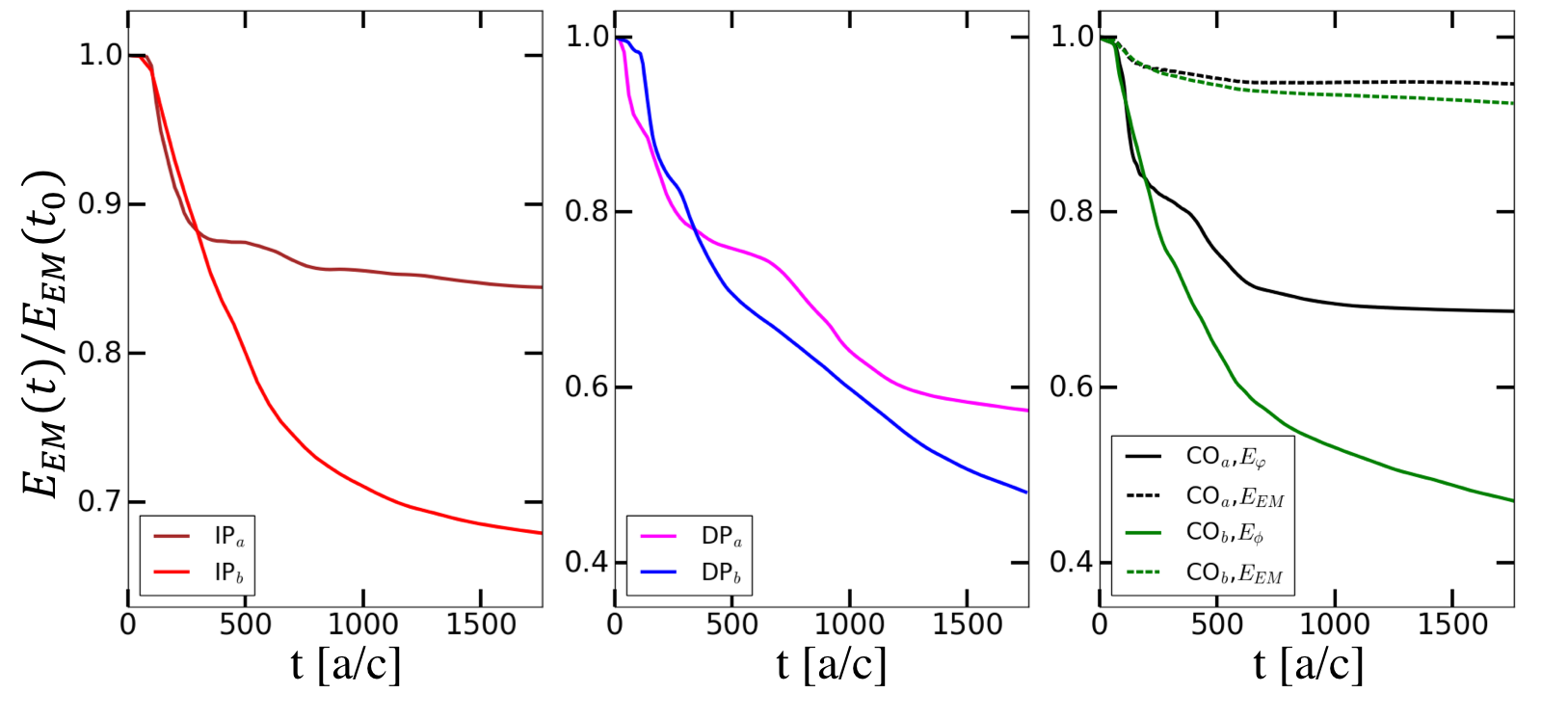}
    \caption{The EM energy dissipation in the three profiles studied. In each profile we show the dissipation in the small and big boxes (sub-indices $a$ and $b$ respectively). We show the total EM energy ($\int(E^2+B^2)dV$) in the box normalized by the value at $t=0$. For the CO case we show also the total energy without the contribution of $B_z$, $\int(E^2+B_\varphi^2)dV$ normalized by its initial value as well. the later is the part that undergoes most of the dissipation in the CO case.}
    \label{fig:dissipate}
\end{figure}

During the linear stage a global current sheet is formed at the edge of the kink mode, in regions where the magnetic field is compressed by the growing amplitude of the mode. Since the volume of the current sheet is small and the magnetic field at the location of the sheet is weak, the dissipated energy is small. 
Figure \ref{fig:dissipate} depicts the value of the EM energy at different times in the three configurations. The initial slow decline in the EM energy evident in all panels marks the dissipation during the linear stage. 

The linear stage ends when the individual kink lobes touch each other and begin to merge. As a result the current sheet, which was confined to the outer edge of the kink mode, extends inwards along the surface of contact between the kink lobes and become prominent 
(figs \ref{fig:Jz_Mosice}, \ref{fig:Jz_Mosice_3D} at $t=200$ [a/c]). The dissipation process in the current sheet 
can be attributed to reconnection of magnetic field lines with varying intersection angles, which is driven by the compression of the merging kink lobes. In the IP and CO profiles the reconnection angle is rather small, while in the DP case the reconnecting fieldlines are close to be anti-parallel. Thus, in the DP case the dissipation rate is faster and the kink evolution differs from the first two cases.

As the merging process progresses, the helical current sheet becomes increasingly thinner until it eventually breaks down to small structures, due to resistive effects.
The sub-structures further break into smaller structures resulting in a turbulent configuration of magnetic field. It gradually fills the entire volume inwards to the current sheet and contributes to the dissipation. The energy which is driven into the current sheet through the mergers of the kink lobes, cascades down to the small scale turbulence, keeping the dissipation rate high. 

Once the merging stops, energy is no longer pumped into the turbulence and the dissipation rate is reduced.
This transition is manifested in fig. \ref{fig:dissipate} as a break in the dissipation rate evident in all panels, occurring at times consistent with the end of the merger episodes.  
In the IP and CO cases the mergers reduce the wave number of the kink mode progressively, until it reaches the minimal value allowed in the box, $n=1/2$. Therefore, the duration of the kink mode inverse cascade depends on the longitudinal size of the box, as more waves are exited in larger boxes. Indeed we see in fig. \ref{fig:dissipate} that the transitions from the fast, merger-driven dissipation to the slower, turbulence dissipation occur at later times in the large boxes. 
In the DP case, the merger is instantaneous and its duration is independent of the box size. The large angles between the reconnecting field lines resulting in pumping of more energy into the current sheets than in the IP and CO cases. This is evident in the higher spike in the electric field seen in fig. \ref{fig:Growth_rate}. As a result, the dissipation rate in the DP case is higher and the total fraction of dissipated EM energy is larger as well (see below).

\subsection{Relaxation}
\begin{figure}
    \begin{center}    \includegraphics[width=1.0\linewidth]{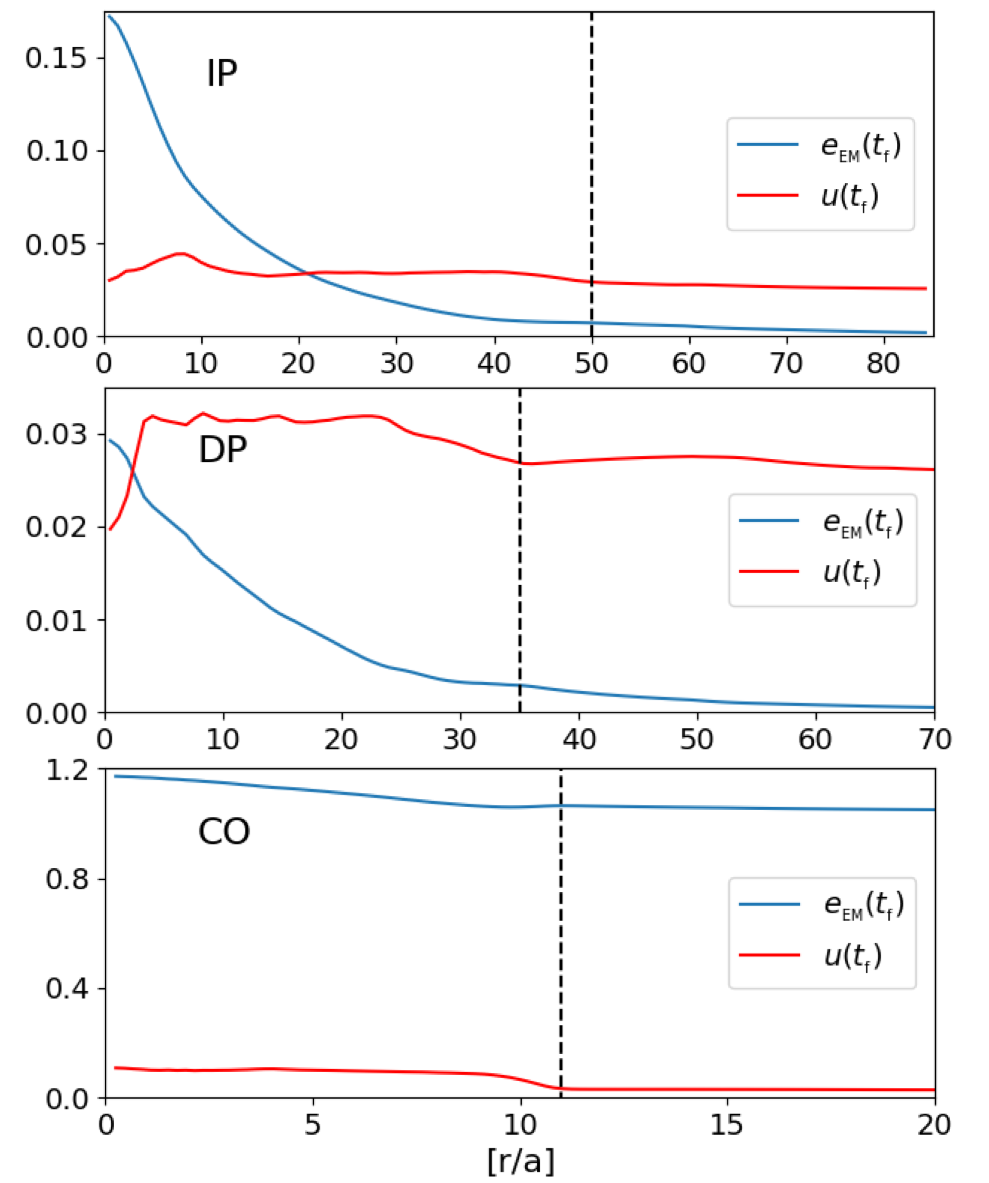}
    \end{center}
    \caption{The EM energy density, $e_{_{\rm EM}}=\frac{1}{8\pi}(E^2+B^2)\simeq \frac{B^2}{8\pi}$ and the 
    thermal energy density, $u=T_{00}-\rho\Gamma^2c^2\simeq 3p$ 
    averaged over $z$ and $\varphi$.
    Shown are the distributions at times $t_f$ from simulations \IPb{} (top), \DPb{} (middle) and \COb{} (bottom). The dashed vertical lines depict the dissipation radius, $R_j$ in the three profiles.
    Though the ratio of magnetic to thermal energy density varies significantly between the profiles, the pressure profile inside $R_j$ is flat, 
    implying that the plasma is dominated by the EM forces and is largely at a force free state.}%
    \label{fig:force-free}%
\end{figure}

The high dissipation rate continues as long as fresh energy is pumped into the turbulence by the inverse cascade of the kink mode. 
Once the KS condition (eq. \ref{eq.KS}) is met, the kink instability relaxes and energy transfer to the turbulence stops. The 
turbulence continues to dissipate the energy contained in them at a slower rate until the system 
reaches a minimal energy state. The magnetic energy configuration at this point is close to a Taylor state, characterized by  
a relatively flat $\alpha$ profile (eq. \ref{eq:alpha}).
During the dissipation process the pressure profile steepens and the pressure gradient becomes of the order of of ${\bf\nabla} B^2/8\pi$. As the configuration approaches the relaxed state, the pressure profile flattens again and the plasma becomes force-free\footnote{In the absence of rotation the transverse force balance equation is $\nabla p+{\rm\bf J}\times{\rm\bf B}=0$. A flat radial pressure profile implies that $({\rm\bf J}\times{\rm\bf B})_r\simeq 0$, thus the plasma is at a force-free state in the transverse direction}, as required by the ideal Taylor state (see eq. \ref{eq.taylor}). Figure \ref{fig:force-free} shows the radial distribution of the EM energy density, $e_{_{\rm EM}}=\frac{1}{8\pi}(E^2+B^2)$, together with  the distribution of the thermal energy density, $u=T_{00}-\rho\Gamma^2c^2$, averaged over $z$ and $\varphi$. In our case Lorentz factors are small and $u\simeq 3p$. It can be seen that although the ratio of EM to thermal energy density varies substantially between the three cases, the final pressure profile is flat and the configuration is force-free.

\begin{figure}
    \centering
    \hspace*{-15pt}
    \includegraphics[width=1.07\linewidth]{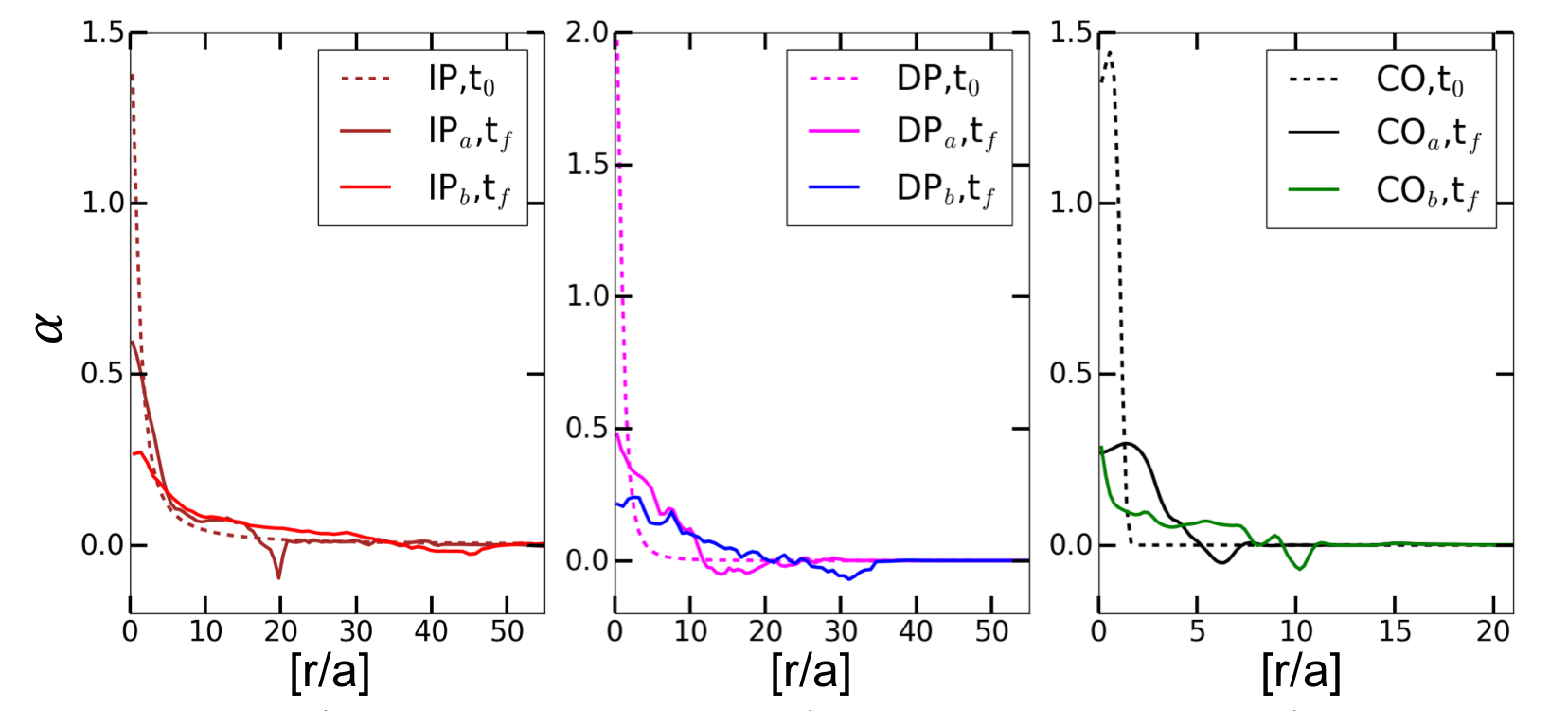}
    \caption{The distribution of $\alpha$, averaged over $z$ and $\phi$ at different times for the three studied profiles. Shown are the profiles at time $t=0$ and at $t=t_f$ in the small and the big boxes (sub-indices $a$ and $b$ respectively). In all cases the distributions in the large boxes are flatter, indicating that the systems are close to a minimal energy state.}
    \label{fig:alpha}
\end{figure}

Figure \ref{fig:alpha} shows the radial profile of $\alpha$ averaged over $z$ and $\varphi$, for all magnetic field profiles and box sizes discussed in this work.
Shown are the initial values (in dashed line) and the values at the end of the simulations. In all large box simulations the $\alpha$ at the core is lower then in the corresponding simulations of small boxes, and it's profile across the box is flatter. This likely occurs since in the large boxes the kink mode has initially a higher wave number, which takes longer to inverse cascade to the lowest $n$. 
As a result the magnetic field distribution has more time to dissipate energy efficiently and thus it can reach a lower energy state.

\begin{figure}
    \centering{
    \hspace*{-15pt}
    \includegraphics[width=0.4\textwidth]{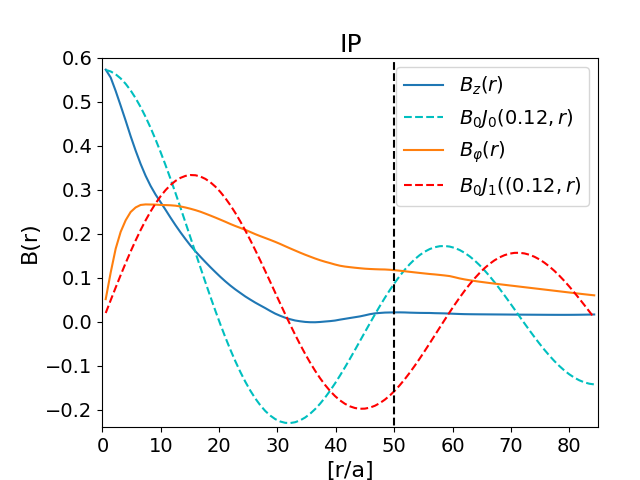}
    \hspace*{-15pt}
    \includegraphics[width=0.4\textwidth]{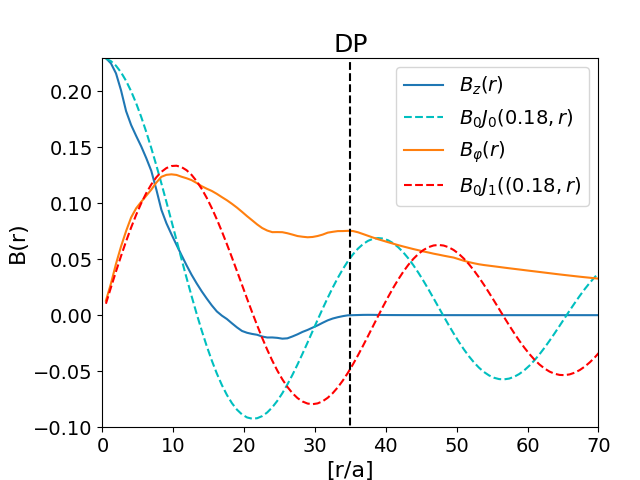}
    \hspace*{-15pt}
    \includegraphics[width=0.4\textwidth]{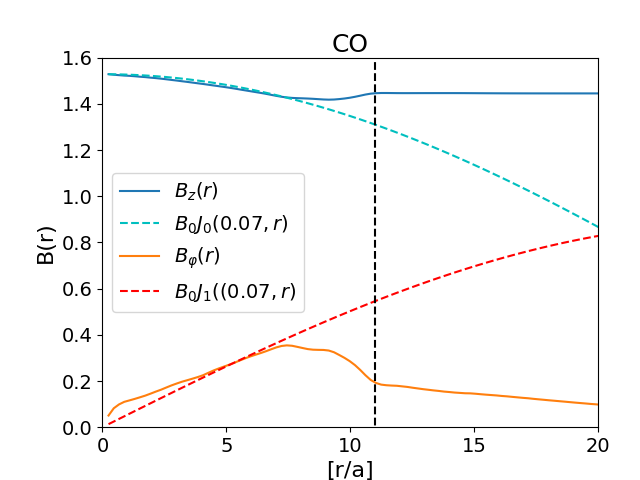}}
    \caption{Fitting $B_0J_0(\alpha r)$ and $B_0J_1(\alpha r)$ to $B_z(r)$ and $B_\varphi(r)$ profiles at the end of each simulation. The best fitted $\alpha$ values are 0.18, 0.07 and 0.12 $[1/a]$, for the DP coronal and IP profiles respectively. The dashed black lines mark the edges of the dissipated regions, $R_j$.}%
    \label{fig:B_Bessel_comp}%
\end{figure}

\begin{figure}
    \centering{
    \includegraphics[width=0.4\textwidth]{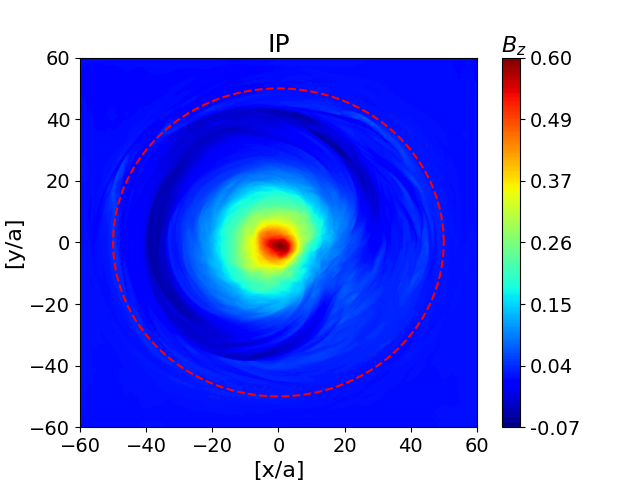}
    \includegraphics[width=0.4\textwidth]{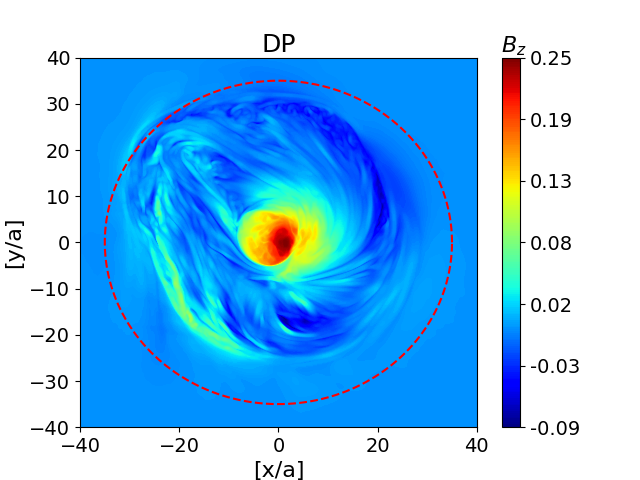} 
    \includegraphics[width=0.4\textwidth]{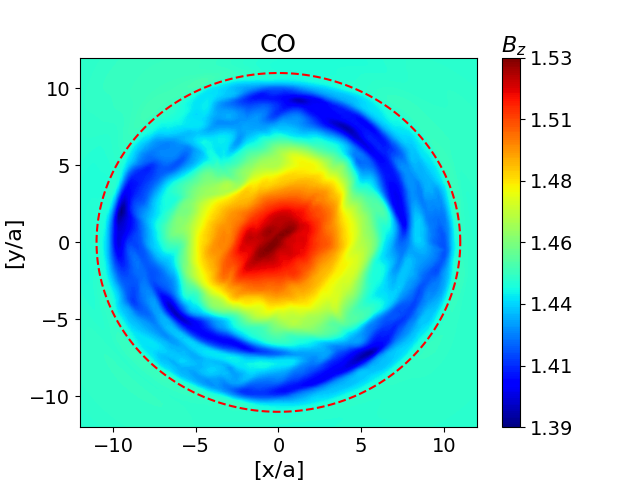}}
    \caption{The value of $B_z$ at at e end of simulations \IPb, \DPb, \COb, shown on a cross-sectional cut in the middle of the computational box. Field reversals are evident in the IP and DP cases but not in the CO case. The dashed red line marks $R_j$ in each case.}%
    \label{fig:Bz2D}%
\end{figure}

The magnetic field configuration relaxes into a Taylor state, which can be represented by 
two Bessel functions of the first kind (see eq. \ref{eq:B_bessel}) with the first zero of $J_0$ falls inside the dissipated region, implying a reversal of $B_z$ close to $R_j$. 
Figure \ref{fig:B_Bessel_comp} shows the magnetic field profiles at the end of runs \IPb{}, \DPb{}, \COb{} averaged over $z$ and $\varphi$. We plot in dotted lines the best fits of the Bessel functions to the configurations. The dashed black lines shows $R_j$. 
It can be seen that in all three cases, the distribution at the central core fits a Taylor profile with the same normalization applied for $B_z$ and $B_\varphi$.
At the outer parts of the dissipated cylinder, the reversal of $B_z$ required by the relaxation criterion is less evident in the IP and DP profiles.
In the IP case this is partly due to the averaging over the azimuthal direction, which washes out the indications of a reversed field.
To demonstrate that we show in fig. \ref{fig:Bz2D} the value of $B_z$ in a cross-sectional cut at the $x$-$y$ plane, in a middle of the box, 
The dashed red line marks $R_j$ in each configuration.
A reversal of the vertical field component is evident in both the IP and the DP cases. No field reversal is seen in the coronal case.

In the CO case the strong magnetic field in the medium, prevents the kink mode from growing to large amplitude with respect to the core cross section, before it breaks down to small scale turbulence. 
Nevertheless, mode merging still occurs in the core, as seen in fig. \ref{fig:Jz_Mosice_3D}, and it likely serves as the energy source for the turbulent dissipation as in the other cases. 
The small amplitude of the kink mode prevents the flip in the direction of $B_z$ from occurring at the outer core part, which is important for obtaining the zero point in $J_0$ seen in the IP and the DP cases. As a result the magnetic field relaxes into a Taylor state with a small $\alpha$, which corresponds to Bessel functions with zero points outside of $R_j$.
The best fitted $\alpha$ values for the three magnetic field configurations in the large boxes are $\alpha=0.18, 0.12, 0.07$ $[1/a]$, for the DP, IP and coronal profiles respectively.

\subsection{Final energy and the minimal energy state}

We find that the dissipation process conserves the total magnetic flux up to $R_j$ and the total helicity with zero gauge, $K$ (eq. \ref{eq:KH}) to $\sim10\%$ in all configurations. A similar fraction of the magnetic energy leaks out through the boundary during the simulation and is likely causing the drop of $K$. Thus, eqs. \ref{eq:KB} and \ref{eq:PsiB} can be used to evaluate the final energy in the box, assuming the system has relaxed to an axially symmetric Taylor state. 
To close the equations we take $R_j$ at the end of each simulation and calculate the values of $\alpha$ and $B_0$ of the corresponding Taylor state. We then compare the EM energy of the Taylor state to the actual EM energy in the box and evaluate how close the system is to a minimal energy. For consistency, we compare the $\alpha R_j$ of the obtained Bessel functions to the theoretical value of a minimal energy state obtained from linear stability analysis (sec. \ref{sec:Emin}). 

\begin{figure}
    \centering{
    \includegraphics[width=0.4\textwidth]{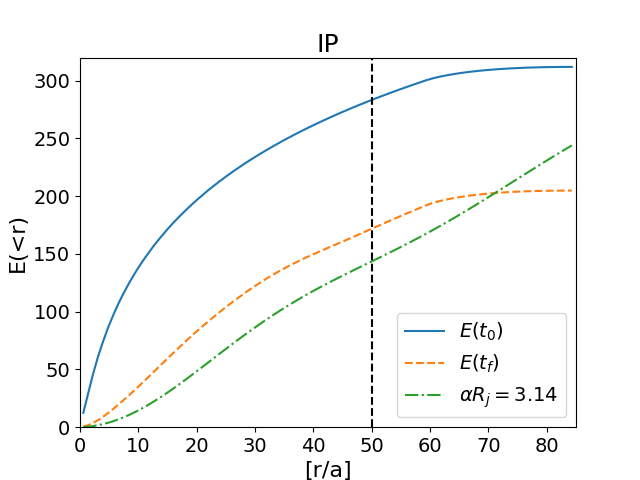}
    \includegraphics[width=0.4\textwidth]{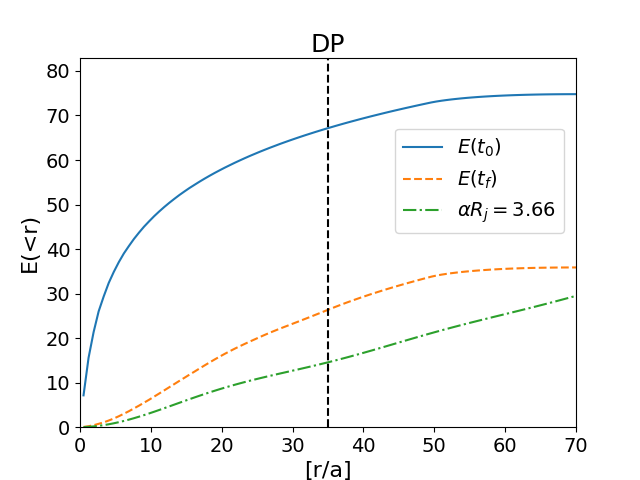} 
    \includegraphics[width=0.4\textwidth]{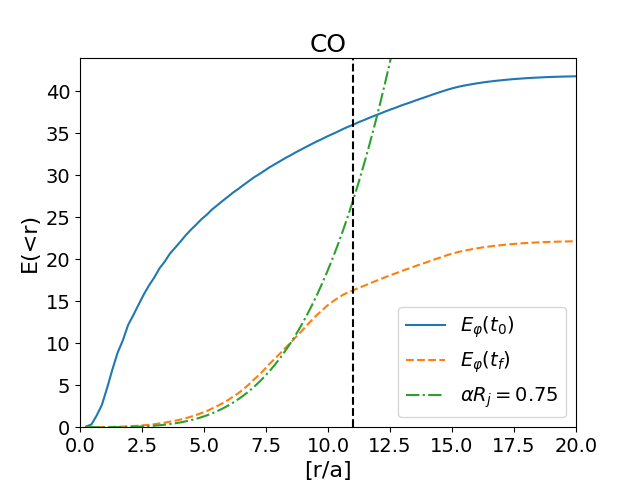}}
    \caption{The initial (solid blue) and final (dashed orange) EM energy in the \IPb, \DPb{} and \COb{} distributions, compared with the estimated energy of the relaxed configuration (dot-dash green). The vertical lines track the radii of the dissipated regions. In the case of CO configuration we show only the energy of $B_\varphi$.}%
    \label{fig:E_comp}%
\end{figure}

Figure \ref{fig:E_comp} shows the total EM energy as a function of $r$ at $t_0$ (blue solid) and $t_{f}$ (orange dashed), for runs \IPb and \DPb. We plot in green (dotted-dash) line the energy distribution of a Taylor state (eq. \ref{eq:B_bessel}) with $B_0$ and $\alpha$ obtained from conservation of $\Psi(R_j)$ and $K(R_j)$ (see Appendix A). The black vertical line shows $R_j$ at each configuration. In the IP case, about $50\%$ of the initial energy is estimated to be available for dissipation, By the end of the simulation $40\%$ of the initial energy has been dissipated, suggesting that the system is close to a minimal energy state. In addition, the obtained $\alpha R_j$ of the Taylor state is very close to the theoretical stability value $\alpha R_j=3.176$  \citep{1962PhRv..128.2016V}. In the DP case $60\%$ of the total EM energy was dissipated by the end of the simulation, where $\sim75\%$ of the total energy up to $R_j$ is estimated to be available for dissipation. Thus, out of the remaining energy about half may still dissipate, implying that the system is still not at a stable state. This result is consistent with the fact that the dissipation in the DP case didn't saturate by the end of the simulation. The obtained value of $\alpha R_j=3.66$ is consistent with the range $3.176\leq \alpha R_j\leq 3.832$ for marginal instability. This  also indicates that the dissipation did not finish evolving to it's minimal energy state. 

In the CO case, the final magnetic field configuration fits a Taylor profile across most of $R_j$  (fig. \ref{fig:B_Bessel_comp}), however with $\alpha R_J\ll3.176$. This manifests the fact that $B_z\gg B_\varphi$ everywhere in the box. By the end of the simulation about $65\%$ of the toroidal field energy inside $R_j$ has been dissipated (fig \ref{fig:E_comp}), which is equivalent to $8\%$ of the total EM energy.
Such a case is inapplicable for the linear stability analysis presented in sec. \ref{sec:theory_relax}, which assumes that the 
first zero of $J_0$ falls inside $R_j$. We are therefore unable to estimate how far is the configuration from the minimal energy state. It is noted that the dissipation by this time did not saturate (fig. \ref{fig:dissipate}).

\section{Astrophysical Implications}
\subsection{Relativistic jets}
Kink instability occurs in narrow plasma columns dominated by toroidal field. Among the systems, which may be affected by such process are collimated relativistic jets. 
A relativistic jet propagating in a medium forms an over pressurized cocoon around it, which applies pressure on the jet and collimates it. At the launching point the jet pressure is much larger than that of the cocoon and the jet expands conically with an initial opening angle $\theta_{0}$. As the jet material expands and accelerates, its pressure drops faster than the pressure of the surroundings until it becomes equal to the cocoon pressure at $\zcol$ and the jet gets collimated.

\begin{figure}
    \centering{
    \includegraphics[width=0.3\textwidth]{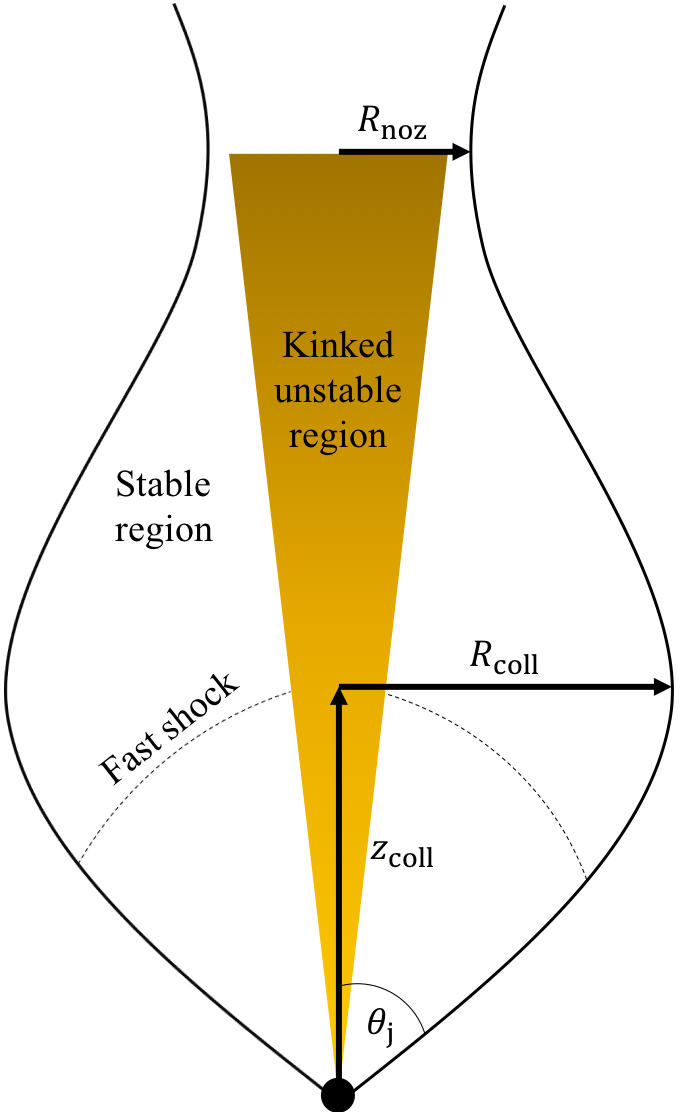}}
    \caption{A sketch of the collimation region of a highly magnetized relativistic jet. The jet is conical up to $z\simeq z_{_{\rm coll}}$, where it's pressure becomes equal to the pressure of the surrounding medium. Above this point the collimated flow is affected by the contracting "hoop stress" of $B_\varphi$ and converges to the axis. Though the converging flow is in strong causal contact it remains stable for kink due to its fast acceleration. At the center of the jet there is a region where the plasma remains sub-superfast and maintains strong lateral causal contact (yellow region). The flow remains in contact with the nozzle and is unable to accelerate efficiently. It can therefore become kink unstable. If the cross section of the unstable region is comparable to $R_{\rm noz}$
    at the nozzle, the converging plasma from the outer, stable parts will interact with it, get shocked and become kinked unstable as well, resulting in an overall dissipation of the jet EM energy.} %
    \label{fig:Jet_coll}%
\end{figure} 

If before the jet plasma reaches $\zcol$ it crosses a fast-magnetosonic surface, the collimation is accompanied by the formation of a weak shock. Downstream of the shock the fluid is sub fast-magnetosonic, and decelerates as it expands until it reaches $\gamma\beta\sim1$ at $\zcol$.
Conservation of magnetic flux implies that the magnetic field value at the collimation point is
\begin{equation}\label{eq:B_coll}
B_{p,c} = B_L\left(\frac{R_L}{\Rcol}\right)^2;~
B_{\phi,c} = B_L\frac{R_L}{\Rcol},
\end{equation}
where $\Rcol=\zcol\theta_{0}$ is the jet cylindrical radius at $\zcol$, $R_L$ is the light cylinder radius, $B_L$ is the magnetic field on that radius and we assume a conical expansion up to $\zcol$. If $\Rcol\gg R_L$ the hoop stress of the toroidal component overcomes the magnetic pressure gradient and the flow converges to the axis \citep{lyub09}. As it contracts, the flow accelerates like \citep{2018MNRAS.480.4948S}  
\begin{equation}\label{eq.gamma_coll}
\gamma=\Rcol/r,
\end{equation} 
where $r$ is the local cylindrical radius, and the magnetic field components in the comoving frame maintain:
\begin{eqnarray}
b_p &\simeq& B_{p,c}\left(\frac{\Rcol}{R}\right)^2\nonumber\\
b_\phi &\simeq& B_{\phi,c}\frac{\Rcol}{R\gamma},
\end{eqnarray}\label{eq:b_comov}
where low cases are used to describe comoving quantities and upper cases for lab frame values. 
The convergence to the axis stops when $b_p\simeq b_\phi$, which by substituting eqs. (\ref{eq:B_coll}), (\ref{eq.gamma_coll}) translates to a nozzle cross sectional radius of
\begin{equation}\label{eq:Rnoz}
    \Rnoz\simeq\sqrt{R_L\Rcol}\simeq\sqrt{R_L\zcol\theta_j}.
\end{equation}

The acceleration of the jet material below the collimation point causes it to loose causal contact with the axis, making the plasma stable to global instabilities such as the kink. As it passes the collimation point and begins to contract, the flow regains causal contact and instabilities can grow. However, the fast acceleration of the flow on the converging flow lines does not allow enough time for the instability to grow in the proper frame, and so the instability grows only linearly with $1/r$ \citep{2018MNRAS.480.4948S}. Thus, ideally kink instability is unlikely to produce strong dissipation in the flow, both below and above the collimation point \citep{2017MNRAS.469.4957B}.

Close to the axis there are field lines with small opening angles, which never loose lateral causal contact.  The flow in this region remains in contact with the nozzle and is unable to accelerate efficiently  below $\zcol$. 
The evolution of the instability in this case is expected to be close to that of a stationary plasma column similar to the ones studied here \citep{2017MNRAS.468.4635S}. \citet{2016MNRAS.456.1739B} obtained a relation for the opening angle of the fieldlines in the unstable region, under the requirement that the plasma on the field lines will be sub fast-magnetosonic and maintain lateral strong causal contact:
\begin{equation}\label{eq:theta_strong}
  \thdiss=
\left\{
\begin{array}{ll}
 \sqrt\frac{R_{\rm L}}{\zcol} & \mbox{, $\zcol<R_{\rm L}\sigma_0^{2/3}$},\\
 \\
 \frac{R_{\rm L}}{\zcol}\sigma_0^{1/3} & \mbox{, $\zcol\geq R_{\rm L}\sigma_0^{2/3}$},  
\end{array}
\right.
\end{equation}

At opening angles $<\thdiss$ the flow is unstable to kink and dissipates its magnetic energy. When it reaches the nozzle it forms an inner core of dissipated plasma. Since most of the toroidal field has dissipated, the core plasma will be less affected by the hoop stress and is not expected to converge to the axis like the outer jet part. Therefore it's opening angle cannot be smaller then $\thdiss$. In fact, it can even be larger due to interaction with material that moves on outer fieldlines, converges onto the dissipated core, get shocked and become kink unstable.   
If the lateral size of the kinked unstable core at the nozzle is comparable to the width of the nozzle (eq. \ref{eq:Rnoz}), most of the plasma passing through the nozzle will get shocked and dissipate its energy. Estimating the radius of the kinked unstable core as $\Rdiss\simeq\thdiss\zcol$ and requiring that at the nozzle $\Rdiss\gtrsim\Rnoz$ we obtain a critical collimation altitude
\begin{equation}\label{eq.z_crit}
    z_{_{\rm crit}}\leq R_L\sigma_0^{2/3}\theta_j^{-1},
\end{equation}
below which
the entire jet material will undergo efficient magnetic dissipation at the nozzle. 
If $\zcol\gg z_{_{\rm crit}}$, the cross sectional radius of the kinked unstable core becomes much smaller then that of the nozzle and most of the jet plasma will pass through the nozzle without interacting with the kinked core and thus may not dissipate its magnetic energy \citep[see e.g.][]{2017MNRAS.469.4957B}. 

In GRBs at the time the jet breaks out of the star, $\zcol\lesssim R_*/10$, where $R_*\simeq 10^{11}$ cm is the stellar radius of the host star. The critical nozzle altitude is, 
\begin{equation}
    z_{_{\rm crit}}\simeq10^{10} R_7\sigma_{_3}^{2/3}\theta_{-1}^{-1} {\rm cm}.
\end{equation}
After the breakout, the cocoon surrounding the jet looses pressure through a rarefaction wave that propagates from the surface inwards towards the collimation point. The wave reaches $\zcol$ a few tens of seconds after the breakout and reduces the cocoon pressure there. As a result the collimation becomes ineffective, leading to a wider nozzle, 
which could stop the magnetic dissipation. This raises an interesting possibility that the observed duration of the prompt GRB emission can be connected with the efficient dissipation of the jet's magnetic energy at the collimation nozzle. Further study of the time evolving conditions at the nozzle before and after the breakout is required to validate this scenario.   

\subsection{Accretion disks}
Kink instability can also play an important role in dissipating magnetic energy of twisted loops above accretion disks. Geometrically thin accretion disks near AGNs can support highly magnetized coronae consisting of small scale magnetic flux tubes \citep[e.g.][]{1979ApJ...229..318G}, which is thought to power a bright compact X-ray source in a "lamppost" or "extended coronae" models \citep{Parfrey2015, Yuan2019}. Flux tubes are twisted by the disk differential rotation and may eventually become kink unstable under the strong confinement from the neighboring vertical field \citep{Yuan2019}. This situation closely resembles the coronal configuration tested in \citep{2013MNRAS.434.3030B} and in this work. Our results imply that the energy of the toroidal magnetic field stored in the loop gets quickly converted into plasma thermal energy via dissipation in multiple current sheets. As we show, the large-scale current sheets break into turbulence that can further dissipate the magnetic energy in a significant fraction of the volume of the disk's corona. Similar flares powered by reconnection in kink unstable overtwisted magnetic loops can happen in magnetospheres of magnetars \citep{2009ApJ...703.1044B}. Simulations of reconnection driven by kink instability in high-sigma plasma in the loop geometrical configuration will be necessary to quantify the dissipation rate and magnetic energy release.

\section{Discussion and Conclusions}

We show that kink instability growing in relativistic magnetized plasma columns can lead to efficient dissipation of the magnetic field, which continues until the configuration relaxes to a state with minimal free energy. In the case of non-rotating columns, this state corresponds to a force-free Taylor state \citep{1974PhRvL..33.1139T}. This, however, requires a global process that efficiently dissipates the magnetic field energy. In this work we show that the mechanism is driven reconnection, induced by the continuous growth of the kink instability (in the non-linear stage) followed by coalescence of the kink mode to lower longitudinal wave-number modes. 

We identify three stages of the dissipation that correspond to three episodes in the evolution of the instability. i) \emph{Non-linear stage}: at the end of the linear stage, the growth of the kink mode saturates. The growing mode shears the magnetic field configuration inwards to the wave front. The twisted column presses against the non-twisted plasma outside forming a prominent helical current sheet at the wave front.
ii) \emph{Mode merging}: as the instability continues to grow the kink lobes, which expand in the longitudinal direction as well, touch each other and begin to merge. The merging process forces the magnetic fields to reconnect at a high rate. 
It also drives small-scale turbulence which breaks the current sheet, mixes the magnetic fields and helps bringing the plasma to the Taylor equilibrium state. iii) \emph{Relaxation}: the growth of the kink instability relaxes once the kink mode reaches the lowest $k$ allowed in the box. The small scale turbulence continues to dissipate the energy contained in them at a slower rate, until the configuration becomes fully relaxed.  

The dissipation rate as well as the total energy dissipated depend on the magnetic field configuration. Configurations in which the pitch is rising have a resonant surface which tends to regulate the dissipation. The mode coalescence is gradual and the wave number decreases progressively to the lowest value. Configurations in which the pitch is decreasing do not have a resonant surface. They are less stable and experience a more instantaneous coalescence of the kink mode into the minimal wave number allowed in the box. As a result, the dissipation rate is higher and the total amount of dissipated energy is larger. 
In our setups by about $40\%$ of the EM energy was dissipated by the end of the simulation in the IP case, compared with $60\%$ of dissipated energy in the DP case (see fig. \ref{fig:E_comp}). We estimated, through linear stability consideration that the available energy for dissipation in these two cases is $50\%$ and $75\%$ for the IP and DP cases respectively.  
The Coronal configuration tested here behaves similar to the IP case, and seem to dissipate a similar fraction of the toroidal field energy.

We find a toroidal field dissipation rate ${\rm d} U_{B\phi}/{{\rm d}t} \approx -0.1 U_{B\phi}/\tau$, where 
\begin{equation}\label{eq:tau_diss}
\tau \approx 20 \pi P_0/v_{A}
\end{equation}
is the growth time of the linear instability. This rate is qualitatively consistent with the measured sideways motion velocity of $0.1$ c, which drives the reconnection in the current sheet at the boundary of the kinked column. 

Our simulations show that the relaxation criterion for kinked induced dissipation is a minimal energy state, close to the Taylor state. Although thermal pressure becomes important during the dissipation we observe it to flatten out during the relaxation resulting in a force-free configuration. 
We therefore conclude that the thermal pressure likely do not play a role in stabilizing the system.
In the cases of monotonic pitch profiles (IP and DP) where internal kink is evolving, the twisting of $B_z$ results in a reversed field at the outer parts of the dissipated region. This allows the system to relax into a Taylor's state, with parameters defined by conditions of marginal stability. We stress that the ideal value was obtained for $m=-1$ kink modes, while the final stage of the evolution is dominated by turbulent dissipation. The connection to the linear stability criterion likely comes from the fact that the energy in the turbulence originate in the inverse cascade of the kink mode, thus they share the same energy reservoir.
In the Coronal case the strong $B_z$ in the ambient medium prevents a field reversal. The topology of the field doesn't change much and the minimal energy state is close to the initial one. During the evolution of the kink instability, the radius of the kinked unstable core is slowly increasing.
The core pushes against the magnetic field in the medium resulting in the growth of instabilities at the boundary, which mixes external matter into the core. The origin and outcome of such mixing needs to be further studied with numerical simulations.

To reach a minimal energy state, the kink mode needs to go through enough merger episodes as it inverse cascade to the lowest wave number allowed in the box, which pumps energy into the turbulence. In our case, this requires a large enough box that will allow for the growth of a kink mode with a large $n$. 
If the computational box is too small, 
the kink instability relaxes before the plasma has time to reach the Taylor state, and the final magnetic energy is higher. Such a situation is seen in the small box simulation of the increasing pitch (\IPa). Turbulence continues to dissipate energy even after the kink instability relaxes, however the dissipation rate is significantly smaller compared to the mode inverse cascade stages. 

Last, we obtained through analytic considerations the conditions in which kink instability can play a significant role in dissipating the magnetic energy in relativistic collimated jets. These conditions need to be verified in global numerical simulations we intend to perform in future work.  

\subsection{Implications for particle acceleration}

In our MHD simulations without explicit resistivity the dissipation happens on the grid scale. The hope is that with sufficient numerical resolution separation of the dissipation scale, e.g. cell size, and the column size is sufficiently large to represent a realistic astrophysical system. To prove this, we checked that our dissipation rates and dissipated energy fractions are converged with numerical resolution (see Appendix B for convergence tests).

To move further, particle-in-cell (PIC) kinetic plasma simulations can provide an insight into how the magnetic dissipation in kink instability results in non-thermal particle acceleration. In Davelaar et al. (submitted to PRL) we perform PIC simulations for the setups studied in this work. We show that if the jet size is sufficiently large, the kink instability grows at a rate very similar to the ideal MHD instability. We also show the current sheets that form in the non-linear phase of the instability accelerate particles in the initially cold plasma to a non-thermal distribution. The current sheets later break into small scale turbulence, similar to what we observe in the MHD simulations, which continues to dissipate magnetic energy into heat. Future PIC simulations with larger scale separation will allow to probe better the interplay between acceleration in turbulence and reconnection \citep{2018ApJ...867L..18Z,2018arXiv180901966Z}.
A complimentary approach for achieving greater scale separation between the jet size and the dissipation scale might be to perform large-scale resistive MHD simulations with resistivity prescription motivated by PIC simulations and trace particles through these simulations \citep{Ripperda2017}.
  
\subsection*{Acknowledgements}
We would like to thank the Flatiron Institute for its generous hospitality and fourteenth summer school of modern astrophysics at MIPT, where part of this work was done. We would also like to thank A. Bhattacharjee,  M. Medvedev, B. Ripperda and A. Tchekhovskoy for stimulating and insightful discussions and useful comments. O.B. and C.S. were funded by an ISF grant 1657/18 and by an ISF (I-CORE) grant 1829/12. J.D. is supported by the ERC Synergy Grant
``BlackHoleCam-Imaging the Event Horizon of Black Holes’' (Grant 610058,
\cite{goddi2017}). The Flatiron Institute is supported by the Simons Foundation.
\software{{\tt PLUTO}\citep{2007ApJS..170..228M,2012ApJS..198....7M}, {\tt python} \citep{travis2007,jarrod2011}, {\tt scipy} \citep{jones2001}, {\tt numpy} \citep{walt2011}, {\tt matplotlib} \citep{hunter2007}, {\tt VisIt} \citep{visit}}.

\onecolumngrid
\begin{appendices}
\section*{appendix a}
A configuration of axially symmetric magnetic field with vanishing $B_r$ on the boundary, evolves while conserving total helicity and  total magnetic flux. 
The helicity of such configuration can be described as:
\begin{equation}\label{eq:HApp}
    H(R)=2\pi L\left[2\int_0^{R}\frac{\Psi(r')}{2\pi}\frac{2I(r')}{r'}dr'+\left.\left(A_z\frac{\Psi}{2\pi}\right)\right|_0^R\right]
\end{equation}
where $\Psi(r)$ is the magnetic flux within radius $r$ defined as
\begin{equation}
    \Psi(r)=2\pi\int_0^{r}B_zr'dr',
\end{equation}
and $I(r)$ is the current within that radius. Taking a gauge $A_z(R)=0$, the second term vanishes and we are left with the first, which we identify as
\begin{equation}\label{eq:KHApp}
    K(R)\equiv 2\int_0^R\frac{\Psi(r')}{2\pi}\frac{2I(r')}{r'}dr'.
\end{equation} 
$K(R_j)$ is largely conserved throughout the evolution of the system.

If the system evolves to a Taylor state, it's magnetic field components can be described by a pair of Bessel functions of the first kind:
\begin{eqnarray}
B_z &=& B_0 J_0(r\alpha)\\
B_\phi &=& B_0 J_1(r\alpha).
\end{eqnarray}
In this case the vector potential can be expressed as $A_\varphi(r)=B_\varphi(r)/\alpha$,  $A_z=B_z(r)/\alpha$ resulting in an helicity
\begin{equation}\label{eq:H2APP}
    H=2\pi L\int_0^{R_j}({\rm\bf A}\cdot{\rm\bf B})rdr=\frac{2\pi L}{\alpha} B_0^2\int_0^{R_j}\left[J_0(\alpha r)^2+J_1(\alpha r)^2\right] rdr.
\end{equation}
This helicity maintains a gauge, $A_z(R_j)=\frac{B_0}{\alpha} J_0(\alpha R_j)$. Substituting that in eq. \ref{eq:HApp}, we can express the helicity in terms of $K$:
\begin{equation}\label{eq:H3App}
    H=2\pi L\left[K(R_j)+\frac{B_0^2}{\alpha}J_0(\alpha R_j)\int_0^{R_j}J_0(\alpha r)rdr\right],
\end{equation}
with
\begin{equation}\label{eq:KBApp}
    K=\frac{B_0^2}{\alpha}\left(\int_0^{R_j}\left[J_0(\alpha r)^2+\newline
    J_1(\alpha r)^2\right]rdr-J_0(\alpha r)\int_0^{R_j}J_0(\alpha r)rdr\right).
\end{equation}
The total flux maintains:
\begin{equation}\label{eq:PsiApp}
    \Psi=2\pi B_0\int_0^{R_j}J_0(\alpha r)rdr,
\end{equation}

To calculate the total energy in the relaxed state we need to obtain thee parameters $B_0, \alpha$ and $R_j$, thus an additional constraint is required in order to close the equations. For example we can take the constraint of $\alpha R_j$ of the minimal energy configuration obtained from linear stability analysis by \citep{1962PhRv..128.2016V}:
\begin{equation}\label{eqApp:alphaR}
    \alpha R_j=3.176
\end{equation}
to get the three unknowns,: 
\begin{eqnarray}
B_0&=&\left(\frac{K}{\tilde\Upsilon}\right)^2\left(\frac{2\pi}{\Psi}J_1(3.176)\cdot 3.176\right)^3\nonumber\\
\alpha&=&\frac{K}{\tilde\Upsilon}\left(\frac{2\pi}{\Psi}J_1(3.176)\cdot 3.176\right)^2\\
R_j&=&\frac{3.176}{\alpha},\nonumber
\end{eqnarray}
with 
\begin{equation}
    \tilde\Upsilon=\int_0^{3.176}\left[J_0(\xi)^2+J_1(\xi)^2\right]\xi d\xi -J_0(3.176)J_1(3.176)\cdot3.176
\end{equation}
is a constant obtained from eq. \ref{eq:KBApp}, and we used the relation $\int_0^RJ_0(r)rdr=RJ_1(R)$.
Alternatively, we can use the $R_j$ of the simulations to extract $B_0$ and $\alpha$ from eqs. \ref{eq:KBApp} and \ref{eq:PsiApp}. This will define the properties of the Taylor state that corresponds to $R_j$ and conserved the helicity and magnetic flux in the box up to $R_j$. We can then compare $\alpha R_j$ to the expected value from linear stability analysis and estimate how close is the distribution to a relaxed, minimal energy state. 

\section*{appendix b}
\begin{figure}[h]
    \centering{
    \includegraphics[width=0.45\textwidth]{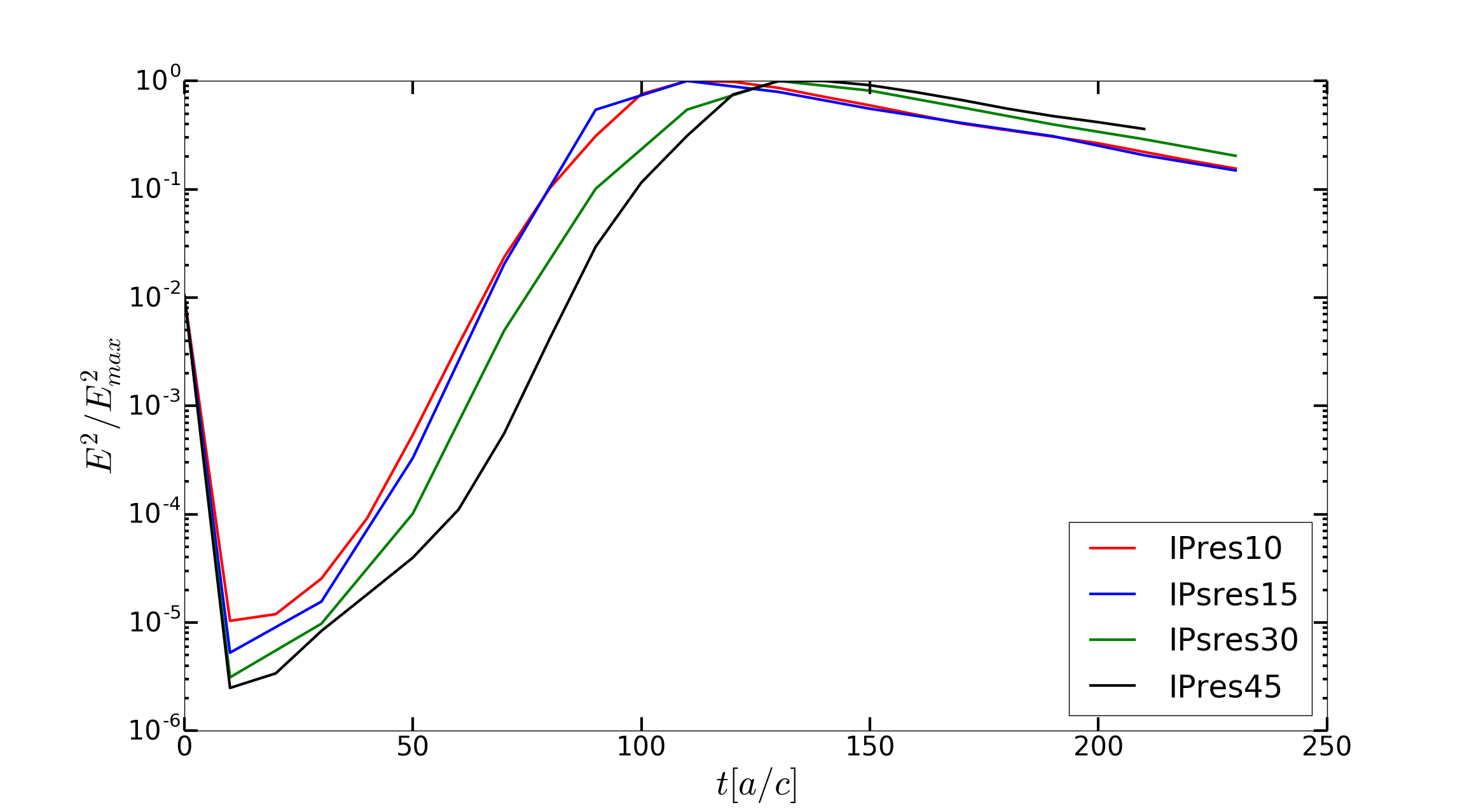}
    \includegraphics[width=0.45\textwidth]{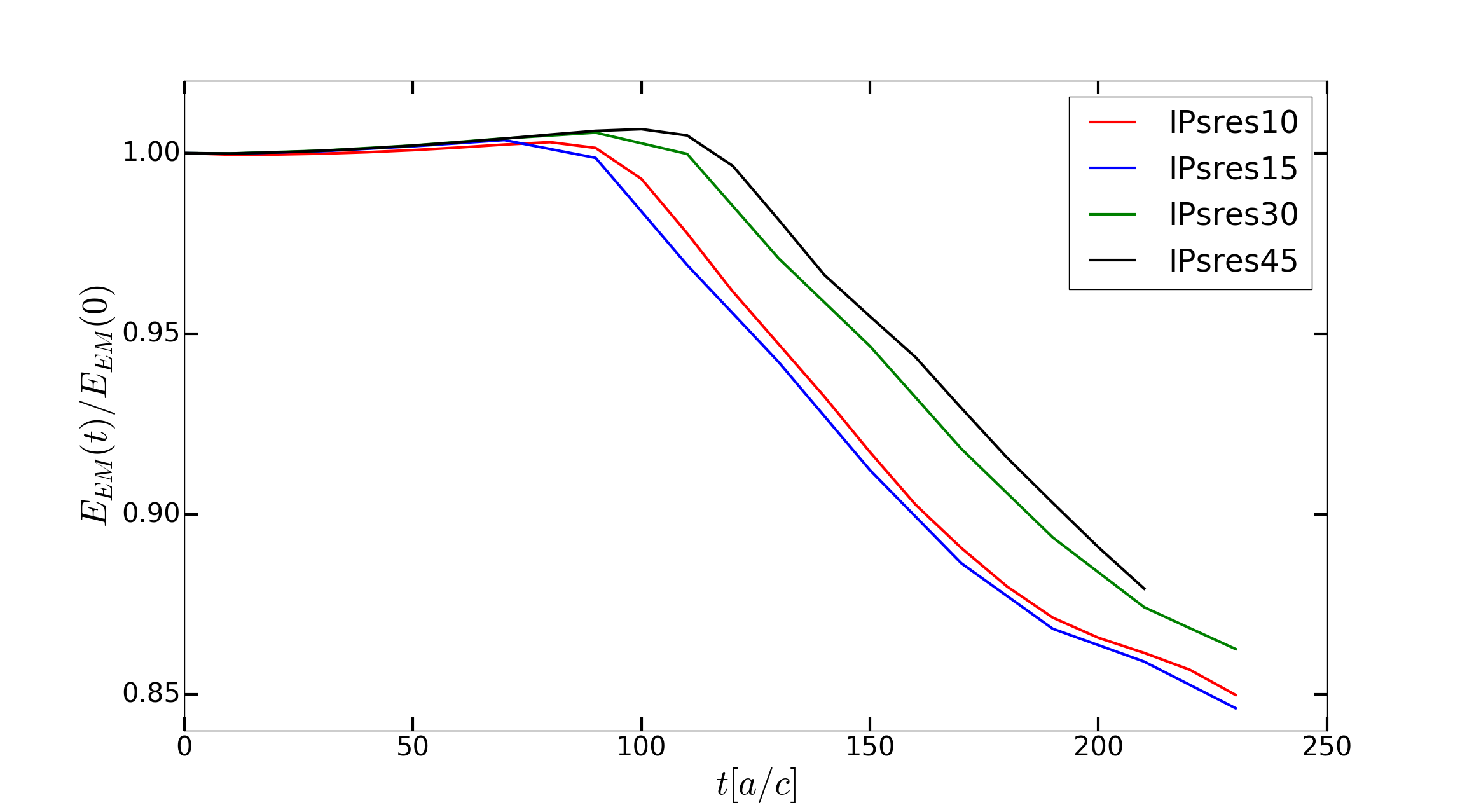}
    }
    \caption{Growth rates of the kink instability, represented by the value of $E^2$ (left) and the EM energy dissipation rates (right) in the IP configuration. Simulations were performed in a box of size $40a\times40a\times20a$ with a resolution of 10, 15, 30 and 45 computational cells per unit length $a$}.  
    \label{fig:App_resolution}%
\end{figure}

Reconnection in ideal MHD simulations is triggered by numerical resistivity. In order to verify that the actual value of the  resistivity doesn't affect the physics of the dissipation process, we examined the dissipation 
during the non-linear stage of the kink instability for different numerical resolution. 
Here we report the tests performed for the IP configuration. We set up the same initial and boundary conditions as in the production runs and compared the evolution for resolutions of 10, 15, 30 and 45 computational cells per unit length $a$. Figure \ref{fig:App_resolution} shows the linear growth rates of the kink instability (left) and the associated EM energy dissipation rates(right). The simulations were made in a smaller box then our production runs, to allow for a manageable run times at high resolutions ($40a\times40a\times20a$). 
The growth rates and the dissipation rates at all four resolutions are almost identical. There is a spread in the peak time of the electric field of $\sim20 a/c$, which corresponds to a similar delay time in the onset of the linear growth. This spread leads to a $2\%$ difference in the dissipated energy at a time of $200 a/c$. We have chosen a resolution of 15 computational cells per unit length $a$ for our production runs. This allows us to run the larger box simulations at a reasonable time and to capture the right physics of the dissipation process.

\end{appendices}


\begin{thebibliography}{}

\bibitem[{Anjiri} {\em et~al.}(2014){Anjiri}, {Mignone}, {Bodo}, and
  {Rossi}]{2014MNRAS.442.2228A}
{Anjiri}, M., {Mignone}, A., {Bodo}, G., and {Rossi}, P. (2014).
\newblock {Linear and non-linear evolution of current-carrying highly
  magnetized jets}.
\newblock {\em \mnras\/}, {\bf 442}(3), 2228--2239.

\bibitem[{Appl}(1996){Appl}]{1996A&A...314..995A}
{Appl}, S. (1996).
\newblock {Instabilities in transmagnetosonic jets.}
\newblock {\em \aap\/}, {\bf 314}, 995--1002.

\bibitem[{Appl} {\em et~al.}(2000){Appl}, {Lery}, and
  {Baty}]{2000A&A...355..818A}
{Appl}, S., {Lery}, T., and {Baty}, H. (2000).
\newblock {Current-driven instabilities in astrophysical jets. Linear
  analysis}.
\newblock {\em \aap\/}, {\bf 355}, 818--828.

\bibitem[{Barniol Duran} {\em et~al.}(2017){Barniol Duran}, {Tchekhovskoy}, and
  {Giannios}]{2017MNRAS.469.4957B}
{Barniol Duran}, R., {Tchekhovskoy}, A., and {Giannios}, D. (2017).
\newblock {Simulations of AGN jets: magnetic kink instability versus conical
  shocks}.
\newblock {\em \mnras\/}, {\bf 469}, 4957--4978.

\bibitem[{Begelman}(1998){Begelman}]{1998ApJ...493..291B}
{Begelman}, M.~C. (1998).
\newblock {Instability of Toroidal Magnetic Field in Jets and Plerions}.
\newblock {\em \apj\/}, {\bf 493}, 291--300.

\bibitem[{Beloborodov}(2009){Beloborodov}]{2009ApJ...703.1044B}
{Beloborodov}, A.~M. (2009).
\newblock {Untwisting Magnetospheres of Neutron Stars}.
\newblock {\em \apj\/}, {\bf 703}(1), 1044--1060.

\bibitem[{Blandford} and {Znajek}(1977){Blandford} and {Znajek}]{bz77}
{Blandford}, R.~D. and {Znajek}, R.~L. (1977).
\newblock {Electromagnetic extraction of energy from Kerr black holes}.
\newblock {\em \mnras\/}, {\bf 179}, 433--456.

\bibitem[{Bodo} {\em et~al.}(2013){Bodo}, {Mamatsashvili}, {Rossi}, and
  {Mignone}]{2013MNRAS.434.3030B}
{Bodo}, G., {Mamatsashvili}, G., {Rossi}, P., and {Mignone}, A. (2013).
\newblock {Linear stability analysis of magnetized relativistic jets: the
  non-rotating case}.
\newblock {\em \mnras\/}, {\bf 434}, 3030--3046.

\bibitem[{Bromberg} and {Tchekhovskoy}(2016){Bromberg} and
  {Tchekhovskoy}]{2016MNRAS.456.1739B}
{Bromberg}, O. and {Tchekhovskoy}, A. (2016).
\newblock {Relativistic MHD simulations of core-collapse GRB jets: 3D
  instabilities and magnetic dissipation}.
\newblock {\em \mnras\/}, {\bf 456}, 1739--1760.

\bibitem[{Browning} {\em et~al.}(2008){Browning}, {Gerrard}, {Hood}, {Kevis},
  and {van der Linden}]{Browning2008}
{Browning}, P.~K., {Gerrard}, C., {Hood}, A.~W., {Kevis}, R., and {van der
  Linden}, R.~A.~M. (2008).
\newblock {Heating the corona by nanoflares: simulations of energy release
  triggered by a kink instability}.
\newblock {\em \aap\/}, {\bf 485}, 837--848.

\bibitem[Childs {\em et~al.}(2005)Childs, Brugger, Bonnell, Meredith, Miller,
  Whitlock, and Max]{visit}
Childs, H., Brugger, E.~S., Bonnell, K.~S., Meredith, J.~S., Miller, M.,
  Whitlock, B.~J., and Max, N. (2005).
\newblock A contract-based system for large data visualization.
\newblock In {\em Proceedings of IEEE Visualization 2005\/}, pages 190--198.

\bibitem[{Freidberg} and {Haas}(1973){Freidberg} and
  {Haas}]{1973PhFl...16.1909F}
{Freidberg}, J.~P. and {Haas}, F.~A. (1973).
\newblock {Kink instabilities in a high-{\ensuremath{\beta}} tokamak}.
\newblock {\em Physics of Fluids\/}, {\bf 16}(11), 1909--1916.

\bibitem[{Galeev} {\em et~al.}(1979){Galeev}, {Rosner}, and
  {Vaiana}]{1979ApJ...229..318G}
{Galeev}, A.~A., {Rosner}, R., and {Vaiana}, G.~S. (1979).
\newblock {Structured coronae of accretion disks.}
\newblock {\em \apj\/}, {\bf 229}, 318--326.

\bibitem[{Goddi} {\em et~al.}(2017){Goddi}, {Falcke}, {Kramer}, {Rezzolla},
  {Brinkerink}, {Bronzwaer}, {Davelaar}, {Deane}, {de Laurentis}, {Desvignes},
  {Eatough}, {Eisenhauer}, {Fraga-Encinas}, {Fromm}, {Gillessen}, {Grenzebach},
  {Issaoun}, {Jan{\ss}en}, {Konoplya}, {Krichbaum}, {Laing}, {Liu}, {Lu},
  {Mizuno}, {Moscibrodzka}, {M{\"u}ller}, {Olivares}, {Pfuhl}, {Porth},
  {Roelofs}, {Ros}, {Schuster}, {Tilanus}, {Torne}, {van Bemmel}, {van
  Langevelde}, {Wex}, {Younsi}, and {Zhidenko}]{goddi2017}
{Goddi}, C., {Falcke}, H., {Kramer}, M., {Rezzolla}, L., {Brinkerink}, C.,
  {Bronzwaer}, T., {Davelaar}, J.~R.~J., {Deane}, R., {de Laurentis}, M.,
  {Desvignes}, G., {Eatough}, R.~P., {Eisenhauer}, F., {Fraga-Encinas}, R.,
  {Fromm}, C.~M., {Gillessen}, S., {Grenzebach}, A., {Issaoun}, S.,
  {Jan{\ss}en}, M., {Konoplya}, R., {Krichbaum}, T.~P., {Laing}, R., {Liu}, K.,
  {Lu}, R.~S., {Mizuno}, Y., {Moscibrodzka}, M., {M{\"u}ller}, C., {Olivares},
  H., {Pfuhl}, O., {Porth}, O., {Roelofs}, F., {Ros}, E., {Schuster}, K.,
  {Tilanus}, R., {Torne}, P., {van Bemmel}, I., {van Langevelde}, H.~J., {Wex},
  N., {Younsi}, Z., and {Zhidenko}, A. (2017).
\newblock {BlackHoleCam: Fundamental physics of the galactic center}.
\newblock {\em International Journal of Modern Physics D\/}, {\bf 26},
  1730001--239.

\bibitem[{Gordovskyy} and {Browning}(2011){Gordovskyy} and
  {Browning}]{2011ApJ...729..101G}
{Gordovskyy}, M. and {Browning}, P.~K. (2011).
\newblock {Particle Acceleration by Magnetic Reconnection in a Twisted Coronal
  Loop}.
\newblock {\em \apj\/}, {\bf 729}, 101.

\bibitem[{Hawley} {\em et~al.}(2015){Hawley}, {Fendt}, {Hardcastle},
  {Nokhrina}, and {Tchekhovskoy}]{2015SSRv..191..441H}
{Hawley}, J.~F., {Fendt}, C., {Hardcastle}, M., {Nokhrina}, E., and
  {Tchekhovskoy}, A. (2015).
\newblock {Disks and Jets. Gravity, Rotation and Magnetic Fields}.
\newblock {\em \ssr\/}, {\bf 191}, 441--469.

\bibitem[{Hood} and {Priest}(1979){Hood} and {Priest}]{1979SoPh...64..303H}
{Hood}, A.~W. and {Priest}, E.~R. (1979).
\newblock {Kink Instability of Solar Coronal Loops as the Cause of Solar
  Flares}.
\newblock {\em \solphys\/}, {\bf 64}(2), 303--321.

\bibitem[{Hunter}(2007){Hunter}]{hunter2007}
{Hunter}, J.~D. (2007).
\newblock {Matplotlib: A 2D Graphics Environment}.
\newblock {\em Computing in Science and Engineering\/}, {\bf 9}, 90--95.

\bibitem[{Istomin} and {Pariev}(1996){Istomin} and
  {Pariev}]{1996MNRAS.281....1I}
{Istomin}, Y.~N. and {Pariev}, V.~I. (1996).
\newblock {Stability of a relativistic rotating electron-positron jet:
  non-axisymmetric perturbations}.
\newblock {\em \mnras\/}, {\bf 281}, 1--26.

\bibitem[Jones {\em et~al.}(2001)Jones, Oliphant, Peterson, {\em
  et~al.}]{jones2001}
Jones, E., Oliphant, T., Peterson, P., {\em et~al.} (2001).
\newblock {SciPy}: Open source scientific tools for {Python}.
\newblock [Online].

\bibitem[{Kadomtsev}(1975){Kadomtsev}]{1975SvJPP...1..389K}
{Kadomtsev}, B.~B. (1975).
\newblock {Disruptive instability in Tokamaks}.
\newblock {\em Soviet Journal of Plasma Physics\/}, {\bf 1}, 710--715.

\bibitem[{Komissarov}(2001){Komissarov}]{kom01}
{Komissarov}, S.~S. (2001).
\newblock {Direct numerical simulations of the Blandford-Znajek effect}.
\newblock {\em \mnras\/}, {\bf 326}, L41--L44.

\bibitem[{Kruskal} and {Tuck}(1958){Kruskal} and {Tuck}]{1958RSPSA.245..222K}
{Kruskal}, M. and {Tuck}, J.~L. (1958).
\newblock {The Instability of a Pinched Fluid with a Longitudinal Magnetic
  Field}.
\newblock {\em Proceedings of the Royal Society of London Series A\/}, {\bf
  245}(1241), 222--237.

\bibitem[{Lery} {\em et~al.}(2000){Lery}, {Baty}, and
  {Appl}]{2000A&A...355.1201L}
{Lery}, T., {Baty}, H., and {Appl}, S. (2000).
\newblock {Current-driven instabilities in astrophysical jets. Non linear
  development}.
\newblock {\em \aap\/}, {\bf 355}, 1201--1208.

\bibitem[{Lyubarskii}(1999){Lyubarskii}]{1999MNRAS.308.1006L}
{Lyubarskii}, Y.~E. (1999).
\newblock {Kink instability of relativistic force-free jets}.
\newblock {\em \mnras\/}, {\bf 308}, 1006--1010.

\bibitem[{Lyubarsky}(2009){Lyubarsky}]{lyub09}
{Lyubarsky}, Y. (2009).
\newblock {Asymptotic Structure of Poynting-Dominated Jets}.
\newblock {\em \apj\/}, {\bf 698}, 1570--1589.

\bibitem[{Mignone} {\em et~al.}(2007){Mignone}, {Bodo}, {Massaglia},
  {Matsakos}, {Tesileanu}, {Zanni}, and {Ferrari}]{2007ApJS..170..228M}
{Mignone}, A., {Bodo}, G., {Massaglia}, S., {Matsakos}, T., {Tesileanu}, O.,
  {Zanni}, C., and {Ferrari}, A. (2007).
\newblock {PLUTO: A Numerical Code for Computational Astrophysics}.
\newblock {\em \apjs\/}, {\bf 170}, 228--242.

\bibitem[{Mignone} {\em et~al.}(2010){Mignone}, {Rossi}, {Bodo}, {Ferrari}, and
  {Massaglia}]{2010MNRAS.402....7M}
{Mignone}, A., {Rossi}, P., {Bodo}, G., {Ferrari}, A., and {Massaglia}, S.
  (2010).
\newblock {High-resolution 3D relativistic MHD simulations of jets}.
\newblock {\em \mnras\/}, {\bf 402}, 7--12.

\bibitem[{Mignone} {\em et~al.}(2012){Mignone}, {Zanni}, {Tzeferacos}, {van
  Straalen}, {Colella}, and {Bodo}]{2012ApJS..198....7M}
{Mignone}, A., {Zanni}, C., {Tzeferacos}, P., {van Straalen}, B., {Colella},
  P., and {Bodo}, G. (2012).
\newblock {The PLUTO Code for Adaptive Mesh Computations in Astrophysical Fluid
  Dynamics}.
\newblock {\em \apjs\/}, {\bf 198}, 7.

\bibitem[{Mignone} {\em et~al.}(2013){Mignone}, {Striani}, {Tavani}, and
  {Ferrari}]{2013MNRAS.436.1102M}
{Mignone}, A., {Striani}, E., {Tavani}, M., and {Ferrari}, A. (2013).
\newblock {Modelling the kinked jet of the Crab nebula}.
\newblock {\em \mnras\/}, {\bf 436}, 1102--1115.

\bibitem[Millman and Aivazis(2011)Millman and Aivazis]{jarrod2011}
Millman, K.~J. and Aivazis, M. (2011).
\newblock Python for scientists and engineers.
\newblock {\em Computing in Science \& Engineering\/}, {\bf 13}(2), 9--12.

\bibitem[{Mizuno} {\em et~al.}(2009){Mizuno}, {Lyubarsky}, {Nishikawa}, and
  {Hardee}]{2009ApJ...700..684M}
{Mizuno}, Y., {Lyubarsky}, Y., {Nishikawa}, K.-I., and {Hardee}, P.~E. (2009).
\newblock {Three-Dimensional Relativistic Magnetohydrodynamic Simulations of
  Current-Driven Instability. I. Instability of a Static Column}.
\newblock {\em \apj\/}, {\bf 700}, 684--693.

\bibitem[{Mizuno} {\em et~al.}(2012){Mizuno}, {Lyubarsky}, {Nishikawa}, and
  {Hardee}]{2012ApJ...757...16M}
{Mizuno}, Y., {Lyubarsky}, Y., {Nishikawa}, K.-I., and {Hardee}, P.~E. (2012).
\newblock {Three-dimensional Relativistic Magnetohydrodynamic Simulations of
  Current-driven Instability. III. Rotating Relativistic Jets}.
\newblock {\em \apj\/}, {\bf 757}, 16.

\bibitem[Oliphant(2007)Oliphant]{travis2007}
Oliphant, T.~E. (2007).
\newblock Python for scientific computing.
\newblock {\em Computing in Science \& Engineering\/}, {\bf 9}(3), 10--20.

\bibitem[{Parfrey} {\em et~al.}(2013){Parfrey}, {Beloborodov}, and
  {Hui}]{2013ApJ...774...92P}
{Parfrey}, K., {Beloborodov}, A.~M., and {Hui}, L. (2013).
\newblock {Dynamics of Strongly Twisted Relativistic Magnetospheres}.
\newblock {\em \apj\/}, {\bf 774}(2), 92.

\bibitem[{Parfrey} {\em et~al.}(2015){Parfrey}, {Giannios}, and
  {Beloborodov}]{Parfrey2015}
{Parfrey}, K., {Giannios}, D., and {Beloborodov}, A.~M. (2015).
\newblock {Black hole jets without large-scale net magnetic flux}.
\newblock {\em \mnras\/}, {\bf 446}, L61--L65.

\bibitem[{Ripperda} {\em et~al.}(2017){Ripperda}, {Porth}, {Xia}, and
  {Keppens}]{Ripperda2017}
{Ripperda}, B., {Porth}, O., {Xia}, C., and {Keppens}, R. (2017).
\newblock {Reconnection and particle acceleration in interacting flux ropes -
  II. 3D effects on test particles in magnetically dominated plasmas}.
\newblock {\em \mnras\/}, {\bf 471}(3), 3465--3482.

\bibitem[{Rosenbluth} {\em et~al.}(1973){Rosenbluth}, {Dagazian}, and
  {Rutherford}]{1973PhFl...16.1894R}
{Rosenbluth}, M.~N., {Dagazian}, R.~Y., and {Rutherford}, P.~H. (1973).
\newblock {Nonlinear properties of the internal m = 1 kink instability in the
  cylindrical tokamak}.
\newblock {\em Physics of Fluids\/}, {\bf 16}, 1894--1902.

\bibitem[{Shafranov}(1956){Shafranov}]{1956AtEnerg.5...38}
{Shafranov}, V.~D. (1956).
\newblock {\em At. Energ.}, {\bf 5}, 38.

\bibitem[{Sobacchi} and {Lyubarsky}(2018){Sobacchi} and
  {Lyubarsky}]{2018MNRAS.480.4948S}
{Sobacchi}, E. and {Lyubarsky}, Y.~E. (2018).
\newblock {Instability induced by recollimation in highly magnetized outflows}.
\newblock {\em \mnras\/}, {\bf 480}, 4948--4954.

\bibitem[{Sobacchi} {\em et~al.}(2017){Sobacchi}, {Lyubarsky}, and
  {Sormani}]{2017MNRAS.468.4635S}
{Sobacchi}, E., {Lyubarsky}, Y.~E., and {Sormani}, M.~C. (2017).
\newblock {Kink instability of force-free jets: a parameter space study}.
\newblock {\em \mnras\/}, {\bf 468}, 4635--4641.

\bibitem[{Taylor}(1974){Taylor}]{1974PhRvL..33.1139T}
{Taylor}, J.~B. (1974).
\newblock {Relaxation of Toroidal Plasma and Generation of Reverse Magnetic
  Fields}.
\newblock {\em Physical Review Letters\/}, {\bf 33}, 1139--1141.

\bibitem[{Taylor}(1986){Taylor}]{1986RvMP...58..741T}
{Taylor}, J.~B. (1986).
\newblock {Relaxation and magnetic reconnection in plasmas}.
\newblock {\em Reviews of Modern Physics\/}, {\bf 58}, 741--763.

\bibitem[{Taylor}(2000){Taylor}]{2000PhPl....7.1623T}
{Taylor}, J.~B. (2000).
\newblock {Relaxation revisited}.
\newblock {\em Physics of Plasmas\/}, {\bf 7}, 1623--1629.

\bibitem[{van der Walt} {\em et~al.}(2011){van der Walt}, {Colbert}, and
  {Varoquaux}]{walt2011}
{van der Walt}, S., {Colbert}, S.~C., and {Varoquaux}, G. (2011).
\newblock {The NumPy Array: A Structure for Efficient Numerical Computation}.
\newblock {\em Computing in Science and Engineering\/}, {\bf 13}(2), 22--30.

\bibitem[{Voslamber} and {Callebaut}(1962){Voslamber} and
  {Callebaut}]{1962PhRv..128.2016V}
{Voslamber}, D. and {Callebaut}, D.~K. (1962).
\newblock {Stability of Force-Free Magnetic Fields}.
\newblock {\em Physical Review\/}, {\bf 128}, 2016--2021.

\bibitem[{Yuan} {\em et~al.}(2019){Yuan}, {Spitkovsky}, {Blandford}, and
  {Wilkins}]{Yuan2019}
{Yuan}, Y., {Spitkovsky}, A., {Blandford}, R.~D., and {Wilkins}, D.~R. (2019).
\newblock {Black hole magnetosphere with small scale flux tubes--II. Stability
  and dynamics}.
\newblock {\em arXiv e-prints\/}.

\bibitem[{Zhdankin} {\em et~al.}(2018a){Zhdankin}, {Uzdensky}, {Werner}, and
  {Begelman}]{2018arXiv180901966Z}
{Zhdankin}, V., {Uzdensky}, D.~A., {Werner}, G.~R., and {Begelman}, M.~C.
  (2018a).
\newblock {Electron and ion energization in relativistic plasma turbulence}.
\newblock {\em arXiv e-prints\/}, page arXiv:1809.01966.

\bibitem[{Zhdankin} {\em et~al.}(2018b){Zhdankin}, {Uzdensky}, {Werner}, and
  {Begelman}]{2018ApJ...867L..18Z}
{Zhdankin}, V., {Uzdensky}, D.~A., {Werner}, G.~R., and {Begelman}, M.~C.
  (2018b).
\newblock {System-size Convergence of Nonthermal Particle Acceleration in
  Relativistic Plasma Turbulence}.
\newblock {\em \apj\/}, {\bf 867}(1), L18.

\end{thebibliography}
\bibliographystyle{natbib.bst}


\end{document}